\documentclass[iop,apj]{emulateapj}
\usepackage{times}
\usepackage{epsfig}
\usepackage{amsmath, amsthm, amssymb}
\usepackage[plainpages=false, colorlinks=true, anchorcolor=blue, linkcolor=blue, citecolor=blue, bookmarks=false]{hyperref}
\usepackage{color}
\usepackage{microtype}
\usepackage{float}

\newcommand{\beq}{\begin{equation}}
\newcommand{\eeq}{\end{equation}}
\newcommand{\bea}{\begin{eqnarray}}
\newcommand{\eea}{\end{eqnarray}}

\slugcomment{Submitted to the Astrophysical Journal}

\begin{document}

\title{On the Impact of Three Dimensions in Simulations of
  Neutrino-Driven Core-Collapse Supernova Explosions} \author{Sean
  M. Couch\altaffilmark{1}}

\affil{Flash Center for Computational Science, Department of Astronomy
  \& Astrophysics, University of Chicago, Chicago, IL, 60637;
  smc@flash.uchicago.edu} \altaffiltext{1}{Hubble Fellow}

\shorttitle{CCSNe DIMENSIONALITY}
\shortauthors{COUCH}

\begin{abstract}

  We present 1D, 2D, and 3D hydrodynamical simulations of
  core-collapse supernovae including a parameterized neutrino heating
  and cooling scheme in order to investigate the critical core
  neutrino luminosity ($L_{\rm crit}$) required for explosion.  In
  contrast to some previous works, we find that 3D simulations explode
  {\it later} than 2D simulations, and that $L_{\rm crit}$ at fixed
  mass accretion rate is somewhat higher in 3D than in 2D.  We find,
  however, that in 2D $L_{\rm crit}$ {\it increases} as the numerical
  resolution of the simulation {\it increases}.  In contrast to some
  previous works, we argue that the average entropy of the gain region
  is in fact not a good indicator of explosion but is rather a
  reflection of the greater mass in the gain region in 2D. We compare
  our simulations to semi-analytic explosion criteria and examine the
  nature of the convective motions in 2D and 3D.  We discuss the
  balance between neutrino-driven-buoyancy and drag forces.  In
  particular, we show that the drag force will be proportional to a
  buoyant plume's surface area while the buoyant force is proportional
  to a plume's volume and, therefore, plumes with greater
  volume-to-surface area ratios will rise more quickly.  We show that
  buoyant plumes in 2D are inherently larger, with greater
  volume-to-surface area ratios, than plumes in 3D.  In the scenario
  that the supernova shock expansion is dominated by neutrino-driven
  buoyancy, this balance between buoyancy and drag forces may explain
  why 3D simulations explode later than 2D simulations and why $L_{\rm
    crit}$ increases with resolution.  Finally, we provide a
  comparison of our results with other calculations in the literature.


\end{abstract}

\section{Introduction}
\label{sec:Intro}

Multidimensional phenomena play a critical role in the core-collapse
supernova (CCSN) mechanism.  Instabilities such as proto-neutron star
convection \citep{Epstein:1979tg,Burrows:1993ki}, neutrino-driven
convection \citep{Herant:1994kk, Burrows:1995gs, Janka:1996wv,
  Murphy:2013eg}, and the standing accretion shock instability
\citep[SASI,][]{Blondin:2003ep,{Blondin:2006dv}} have abetted
explosions in 2D \citep{Marek:2009kc,Muller:2012gd,{Muller:2012kq}}
for progenitors that refuse to explode in spherical symmetry
\citep{Rampp:2000gd,Liebendorfer:2001fl,Thompson:2003kn,{Suwa:2010wp}}.
The additional degrees of freedom afforded by multiple dimensions can
also increase the dwell time of matter in the post-shock region where
the accreting matter experiences a net gain of neutrino energy
resulting in an increased efficiency of neutrino heating.  Taken
together, these multidimensional effects lower the critical neutrino
luminosity threshold for which explosions are obtained when comparing
2D to 1D \citep{Murphy:2008ij}.  In spherical symmetry, the radial
stability of the supernova shock has been studied in detail
\citep{Burrows:1993ft,
  {Yamasaki:2007kz},Pejcha:2012cw,Fernandez:2012kg}.  The current
state of the research into the CCSN mechanism has been reviewed by
\citet{Janka:2012jp}, \citet{Burrows:2013hp}, and \citet{Janka:2012cb}.

We lack a first principles understanding of how the afore-mentioned
multidimensional effects result in a lowered critical threshold for
explosion in 2D, however, it is attractive to extrapolate the trend to
3D and postulate that the threshold for explosion will be reduced even
further, perhaps permitting energetic explosions for the vast majority
of progenitors. \citet{Nordhaus:2010ct} found precisely this using a
simplified prescription for neutrino heating and cooling and
deleptonization.  Nordhaus et al. found that the critical luminosity
in 3D was reduced by 15\% - 25\% as compared to 2D simulations for a
15 $M_\sun$ progenitor.  Using a similar parametric approach,
\citet{Hanke:2012dx} attempted to reproduce the results of Nordhaus et
al. for both the 15 $M_\sun$ progenitor and a 11.2 $M_\sun$
progenitor.  Contrary to the findings of Nordhaus et al., Hanke et
al. find that there is little difference between the critical
luminosities in 2D and 3D, while recovering the result that the
critical luminosity in 2D is significantly lower than that in 1D
\citep[][]{Murphy:2008ij}. While differences exist between the
approaches of Nordhaus et al. and Hanke et al., the exact cause of
their disparate results is unclear.  \citet{Burrows:2012gc} report
that the results of Nordhaus et al. were beset by inaccuracies in the
gravity solver in CASTRO that have since been corrected, and very
recently \citet{Dolence:2013iw} present new 3D CASTRO simulations
showing faster shock expansion in 3D than in 2D. The
important question of whether the threshold for explosion is lower in
three dimensions is, thus, still an open one.

We are on the precipice of achieving 3D simulations of core-collapse
supernovae with full spectral neutrino transport and adequate
resolution, though a number of 3D CCSN simulations employing various
approximations have already been accomplished. The 3D simulations
to-date have been facilitated by approximations, typically to the
neutrino transport, or by low resolution.  Here we briefly mention a
non-exhaustive list of 3D results relevant to CCSNe extant in the
literature.  The 3D calculations of \citet{Mueller:1997wg} and
\citet{Khokhlov:1999hp} did not include the effects of neutrinos and
excised the PNS from the domain.  A number of early 3D simulations
utilized the smoothed-particle hydrodynamics approximation \citep{
  Fryer:2002hq, Fryer:2004kv, Fryer:2007jd}, which has certain
disadvantages over grid-based codes \citep[see, e.g.,][]{Plewa:2001wx,
  Agertz:2007gd, McNally:2012dj, Sijacki:2012cv}.  Studies of the SASI
have made various approximations to achieve 3D simulations
\citep{Blondin:2007fk, Iwakami:2008dw, Iwakami:2009er,
  Fernandez:2010ko}.  Simulations neglecting the effects of neutrinos
and employing a simplified equation of state (EOS) have been used to
study the amplification of magnetic fields in 3D \citep{Endeve:2010gq,
  Endeve:2012ht}.  \citet{Hammer:2010di}, using a neutrino
``lightbulb'' scheme \citep{Scheck:2006iz}, followed the evolution of
the supernova explosion all the way through the envelope of the
progenitor in 3D and examined the asymmetric development of
instabilities.  Some studies have focussed on highly-magnetized
progenitors with approximate (or no) treatments of neutrino effects
\citep{Kuroda:2010dg, Winteler:2012fv}.  The hydrodynamical kick
imparted to the PNS has been studied in 3D by
\citet{Wongwathanarat:2010ch,Wongwathanarat:2013fx}.  The rotational
stability of the PNS has been explored in 3D by \citet{Ott:2005gx}.
Many 3D calculations have focussed on the emergent gravitational wave
signal from CCSNe \citep{Ott:2007cm, Ott:2011jk, Ott:2012ib,
  Scheidegger:2008ds, Scheidegger:2010fe,Scheidegger:2010bk,
  Kotake:2009eq,Muller:2012bz}.  \citet{Ott:2013gz} examined the
development of the SASI in 3D general relativistic simulations with
neutrino leakage.  Some studies have focused also on the neutrino signal
from 3D CCSNe \citep{ {Lund:2012kd},Ott:2012ib}.
\citet{Takiwaki:2012ck} present low-resolution 3D simulations with
spectral neutrino transport using the isotropic diffusion source
approximation \citep[IDSA,][]{Liebendorfer:2009kw}. And recently
fully-3D Boltzmann transport for neutrinos has been developed by
\citet{Sumiyoshi:2012gb}.

In this paper, we describe our multidimensional study of
neutrino-driven CCSN explosions using a parameterization of the
neutrino effects similar to that of \citet{Nordhaus:2010ct} and
\citet{Hanke:2012dx}.  We find that the delay time until explosion for
a given neutrino luminosity is greater in 3D than in 2D, i.e., $L_{\rm
  crit}$ is greater in 3D than in 2D. In Section \ref{sec:Method} we
describe our computational approach.  In Section \ref{sec:Results}, we
present our results. In Section \ref{sec:interpret} we discuss the
dependence of the critical neutrino luminosity on dimensionality and
resolution and suggest that our results can be understood by
considering the balance between buoyant and drag forces acting on
neutrino-driven bubbles.  We also examine the difference in the
character of the shock motion between 2D and 3D.  We demonstrate the
resolution dependence and convergence of our results in Section
\ref{sec:resolution}. In Section \ref{sec:Discussion} we discuss the
implications of our results and we conclude in Section
\ref{sec:Conclusions}.

\begin{figure}
\centering
\includegraphics[width=3.35in,trim=1in .75in 1in .65in,clip]{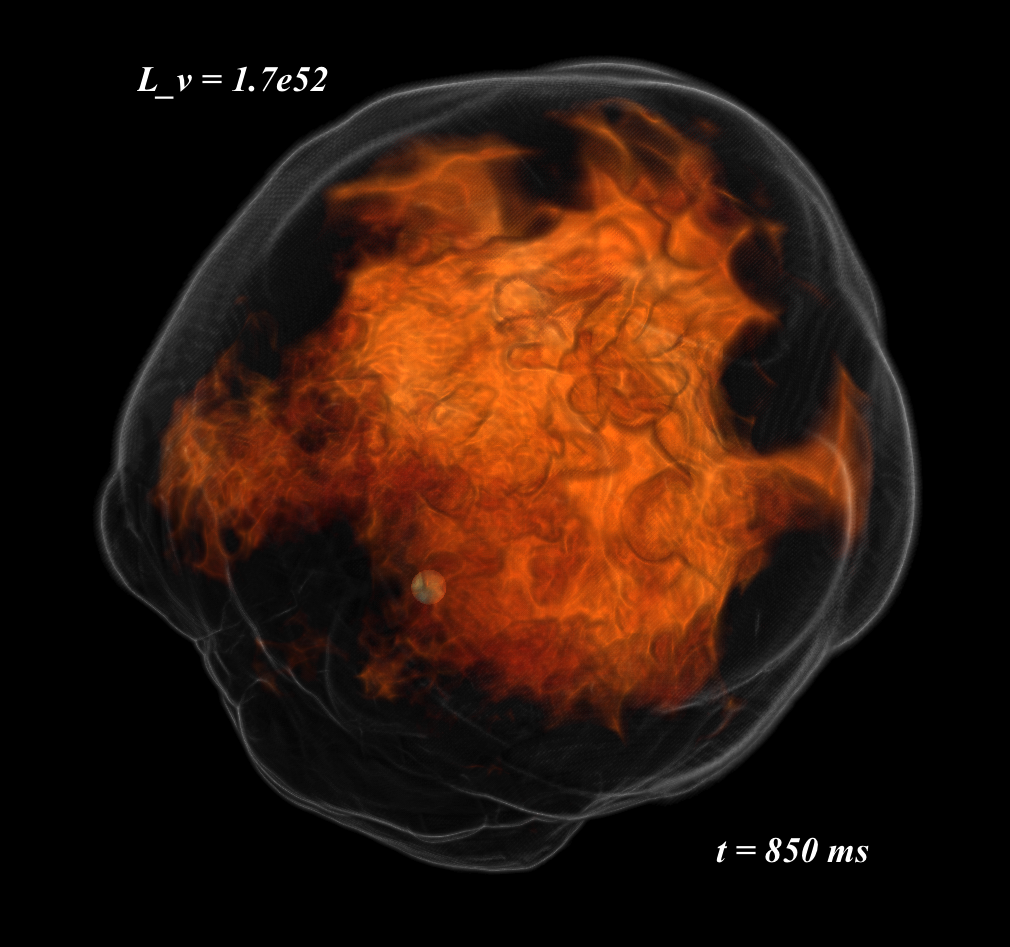}
\caption{Volume rendering of entropy for values above 14 $k_B$
  baryon$^{-1}$ for the 3D model with $L_{\nu_e,52}=1.7$ at 850 ms
  post-bounce.  The shock front is also volume-rendered in gray-scale.
  The blue sphere is a isodensity contour marking the edge of the
  proto-neutron star.  The explosion develops in a non-symmetric
  manner with the PNS recoiling in the direction opposite the dominant
  direction of the explosion.  The high-entropy buoyant plumes display
  a great deal of small-scale structure in 3D.}
\label{fig:volRender}
\end{figure}

\section{Numerical Method}
\label{sec:Method}

Our numerical simulation approach is similar to that described by
\citet{Couch:2013df}. We solve the Eulerian equations of hydrodynamics,
\begin{align}
\frac{\partial \rho}{\partial t} + \pmb{\nabla} \cdot (\rho \pmb{ v}) &= 0,\label{eq:massCons} \\ 
\frac{\partial \rho {\pmb v}}{\partial t} + \pmb{\nabla} \cdot (\rho \pmb{ v v}) + \pmb{\nabla} P &= -\rho \pmb{\nabla} \Phi, \\
\frac{\partial \rho E}{\partial t} + \pmb{\nabla} \cdot [(\rho E + P){\pmb v}] &= \rho {\pmb v} \cdot \pmb{\nabla} \Phi + \rho (\mathcal{H-C}), \label{eq:enerCons}
\end{align}
where $\rho$ is the mass density, $\pmb v$ the velocity vector, $P$
the pressure, $\Phi$ the gravitational potential, $E$ the total
specific energy, $\mathcal{H}$ is the specific neutrino heating, and
$\mathcal{C}$ is the specific neutrino cooling. We use the
directionally-unsplit hydrodynamics solver provided by the FLASH
simulation framework \citep[][, Lee et al., in prep.]{Dubey:2009wz} to
solve equations (\ref{eq:massCons}) - (\ref{eq:enerCons}).  We use
third-order piecewise-parabolic spatial reconstruction
\citep[PPM,][]{{Colella:1984cg}} and a hybrid Riemann solver that
uses the HLLE solver inside of shocks and the HLLC solver in smooth
flow.  We use a ``hybrid'' slope limiter that applies the monotonized
central ({\tt mc}) limiter to linear wave families and the more
diffusive {\tt minmod} limiter to non-linear, self-steepening wave
families.  We use a monopole approximation to calculate the
self-gravity of the flow. To facilitate comparison with
\citet{Nordhaus:2010ct} and \citet{Hanke:2012dx}, we use the
\citet{Shen:1998kx} equation of state (EOS), as implemented by
\citet{OConnor:2010bi}.

Our approach for neutrino heating and cooling is that described by
\citet{Murphy:2008ij}, the same approach used by
\citet{Nordhaus:2010ct} and \citet{Hanke:2012dx}.  The neutrino
heating and cooling are given by,
\begin{align}
\mathcal{H} = 1.544\times10^{20} \left (\frac{L_{\nu_e}}{10^{52}\ {\rm erg\ s^{-1}}} \right ) \left (\frac{T_{\nu_e}}{4\ {\rm MeV}} \right )^2 \nonumber \\
\times \left (\frac{100\ \rm km}{r} \right )^2 (Y_p + Y_n) e^{-\tau_{\nu_e}}\ \left [ \rm{\frac{erg}{g \cdot s}} \right ],\label{eq:heat}
\end{align}
and 
\beq \mathcal{C} = 1.399\times10^{20} \left (\frac{T}{2\ \rm MeV}
\right )^6 (Y_p + Y_n) e^{-\tau_{\nu_e}}\ \left [ \rm{
    \frac{erg}{g\cdot s}} \right ],\label{eq:cool} \eeq 
where
$L_{\nu_e}$ is the electron neutrino luminosity (note it is assumed
that $L_{\bar \nu_e}=L_{\nu_e}$), $T_{\nu_e}$ is the electron neutrino
temperature, $r$ is the spherical radius, $(Y_p + Y_n)$ is the sum of
the neutron and proton number fractions, $\tau_{\nu_e}$ is the
electron neutrino optical depth, and $T$ is the matter temperature.
In all of our simulations, $T_{\nu_e}$ is set to 4 MeV.  We
approximate the neutrino optical depth by a piecewise fit based only
on the density \citep[see][]{Couch:2013df}.  This avoids the need to
calculate radial integrals for the optical depth and is justified
because the factor $e^{-\tau_{\nu_e}}$ is included in equations
(\ref{eq:heat}) \& (\ref{eq:cool}) only as a cutoff to the neutrino
source terms at high-densities.  Differences in the implementation of
this cutoff result in different normalizations for the critical
luminosity curves (J. Murphy, private communication).
\citet{Hanke:2012dx} chose to adjust the neutrino opacities used so
that their 1D critical curves matched those of Nordhaus et al.  We
have not.

We follow the approach proposed by \citet{Liebendorfer:2005ft} for
following the evolution of the electron fraction, $Y_e$.  In this
approach, calibrated with 1D Boltzmann transport simulations, $Y_e$ is
dependent only on density.  This is strictly only applicable during
the pre-bounce collapse phase, however, we continue to use the
density-dependent electron fraction approach post-bounce, as done by
\citet{Nordhaus:2010ct}.  We have found that neglecting any changes in
$Y_e$ post-bounce results in substantially earlier explosions for a
given neutrino luminosity.  We do not include the entropy changes due
to deleptonization given in \citet{Liebendorfer:2005ft}.

We use 1D spherical, 2D cylindrical, and 3D Cartesian geometries with
adaptive mesh refinement (AMR) as implemented in FLASH via PARAMESH
\citep[v.4-dev,][]{{MacNeice:2000fc}}.  For this study we use a
fiducial resolution at the maximum refinement level of 0.7 km in each
direction.  We limit the maximum refinement level with radius such
that a pseudo-logarithmic radial grid spacing is obtained.  Our
refinement limiter takes the form \beq \Delta x^\ell_i > \eta
r, \label{eq:refinement} \eeq where $\Delta x^\ell_i$ is the grid
spacing in the $i$-direction at refinement level $\ell$, $r$ is the
spherical radius, and $\eta$ is a parameter that sets the effective
angular spacing.  If equation (\ref{eq:refinement}) is not satisfied
by a given AMR block, further refinement of that block is prohibited.
For our fiducial resolution we set the finest grid spacing to 0.7 km
and $\eta=1.25\%$, resulting in an effective ``angular'' resolution of
$0\fdg54$.  In 1D, the simulated domain spans 0 km to 5000 km, in 2D
the domain is 0 km to 5000 km in cylindrical radius, $R$, and -5000 km
to 5000 km in $z$, and in 3D the domain is -5000 km to 5000 km in each
Cartesian dimension.  At the outer spatial limits of the domain, we
set boundary conditions that apply power-law profiles to density and
velocity that approximate the stellar envelope outside the domain.
Such boundary conditions are critically important to the results of
the present study as simple ``outflow'' boundary conditions
overestimate the mass accretion rate at late times, altering the
explosion time for near-critical luminosities.  This is because
``outflow'' boundary conditions enforce a zero-gradient condition for
the flow variables which mimics a flat density (etc.) profile outside
the simulation domain artificially enhancing the mass flux into the
domain from the boundary.

We use the 15 $M_\sun$ progenitor of \citet{Woosley:1995jn} in all of
our simulations.

\section{Results: Explosion Times}
\label{sec:Results}


\begin{deluxetable}{ccccccc}
\tablecolumns{7}
\tabletypesize{\scriptsize}
\tablecaption{
Explosion times and accretion rates at time of explosion.
\label{table:results}
}
\tablewidth{0pt}
\tablehead{
\colhead{} &
\multicolumn{2}{c}{0.5 km} &
\multicolumn{2}{c}{0.7 km} &
\multicolumn{2}{c}{0.7 km} \\
\colhead{$L_{\nu_e}$\tablenotemark{a}} &
\colhead{$t_{\textrm{exp}}$\tablenotemark{b}} &
\colhead{$\dot{M}_{\textrm{exp}}$\tablenotemark{c}} &
\colhead{$t_{\textrm{exp}}$} &
\colhead{$\dot{M}_{\textrm{exp}}$} &
\colhead{$t_{\textrm{exp}}$} &
\colhead{$\dot{M}_{\textrm{exp}}$} \\
\colhead{($10^{52}$ erg/s)} &
\colhead{(ms)} &
\colhead{($M_{\odot}$/s)} &
\colhead{(ms)} &
\colhead{($M_{\odot}$/s)} &
\colhead{(ms)} &
\colhead{($M_{\odot}$/s)} }
\startdata 
\cutinhead{1D} 
2.0 &    \nodata   &  \nodata  &  \nodata  &  \nodata    &        &     \\
2.1 &              &           &           &             &        &     \\
2.2 &    \nodata   &  \nodata  &  \nodata  &  \nodata    &        &     \\
2.3 &    943       &  0.153    &  822         &   0.170          &        &     \\
2.4 &    538       &  0.221    &  554         &   0.221          &        &      \\
2.5 &    380       &  0.262    &  389         &   0.262          &        &      \\
2.7 &    216       &  0.310    &  212        &   0.310          &        &      \\
2.9 &    200       &  0.314    &  197         &   0.317          &        &      \\
\cutinhead{ \hspace{.6in} 2D \hspace{.7in} 2D \hspace{.6in} 3D \hspace{.5in}}
1.7 &    713          &  0.190         &  388       &  0.260     &   821      &  0.175   \\
1.8 &    490          &  0.233         &  309       &  0.274     &         &     \\
1.9 &    313          &  0.278         &  291       &  0.284     &   403      &  0.261   \\
2.0 &                 &                &  263       &  0.294     &        &     \\
2.1 &    247          &  0.298         &  222      &   0.313     &   238     &   0.302  \\
\enddata
\tablenotetext{a}{Electron-neutrino luminosity.}
\tablenotetext{b}{Time after bounce of onset of explosion. 
A ``...'' symbol indicates 
that the model does not explode during the simulated period of evolution.}
\tablenotetext{c}{Mass accretion rate at onset of explosion.}
\end{deluxetable}

\begin{figure}
\centering
\includegraphics[width=3.5in,trim= 0in 0in 0.25in 0in, clip]{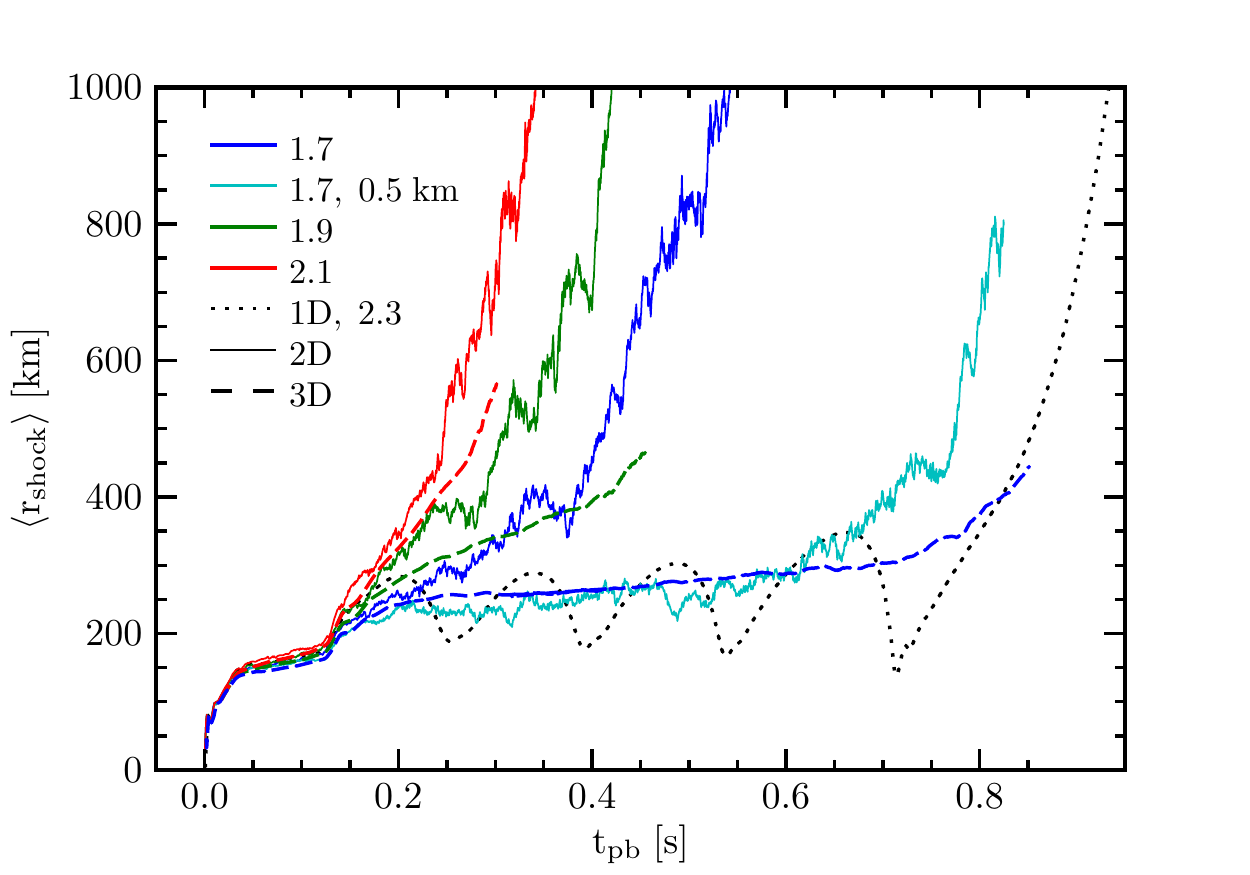}
\caption{Average shock radii as a function of time relative to bounce
  for three neutrino luminosities in 2D and 3D.  Also shown for
  comparison is the shock radius from the 1D simulation with
  $L_{\nu_e,52}=2.3$.  Universally, the shock expands more rapidly in 2D
  than in 3D.  Increasing the resolution in 2D delays explosion, as
  shown by the cyan curve.}
\label{fig:rshock}
\end{figure}

\begin{figure*}
\centering
\begin{tabular}{cc}
\includegraphics[width=3.45in,trim= 0.25in 0in 0.7in 0.5in,clip]{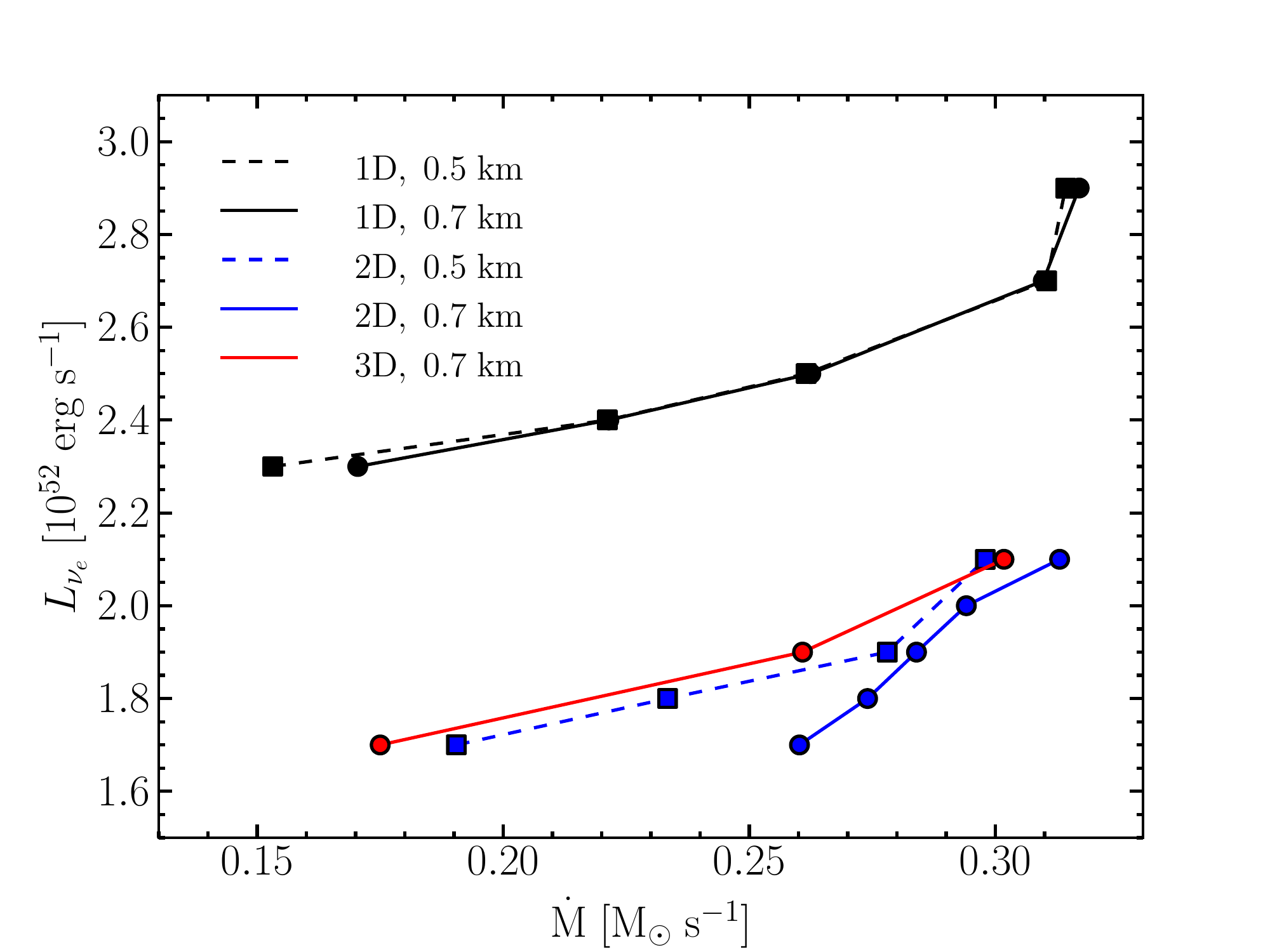} &
\includegraphics[width=3.45in,trim= 0.25in 0in 0.7in 0.5in,clip]{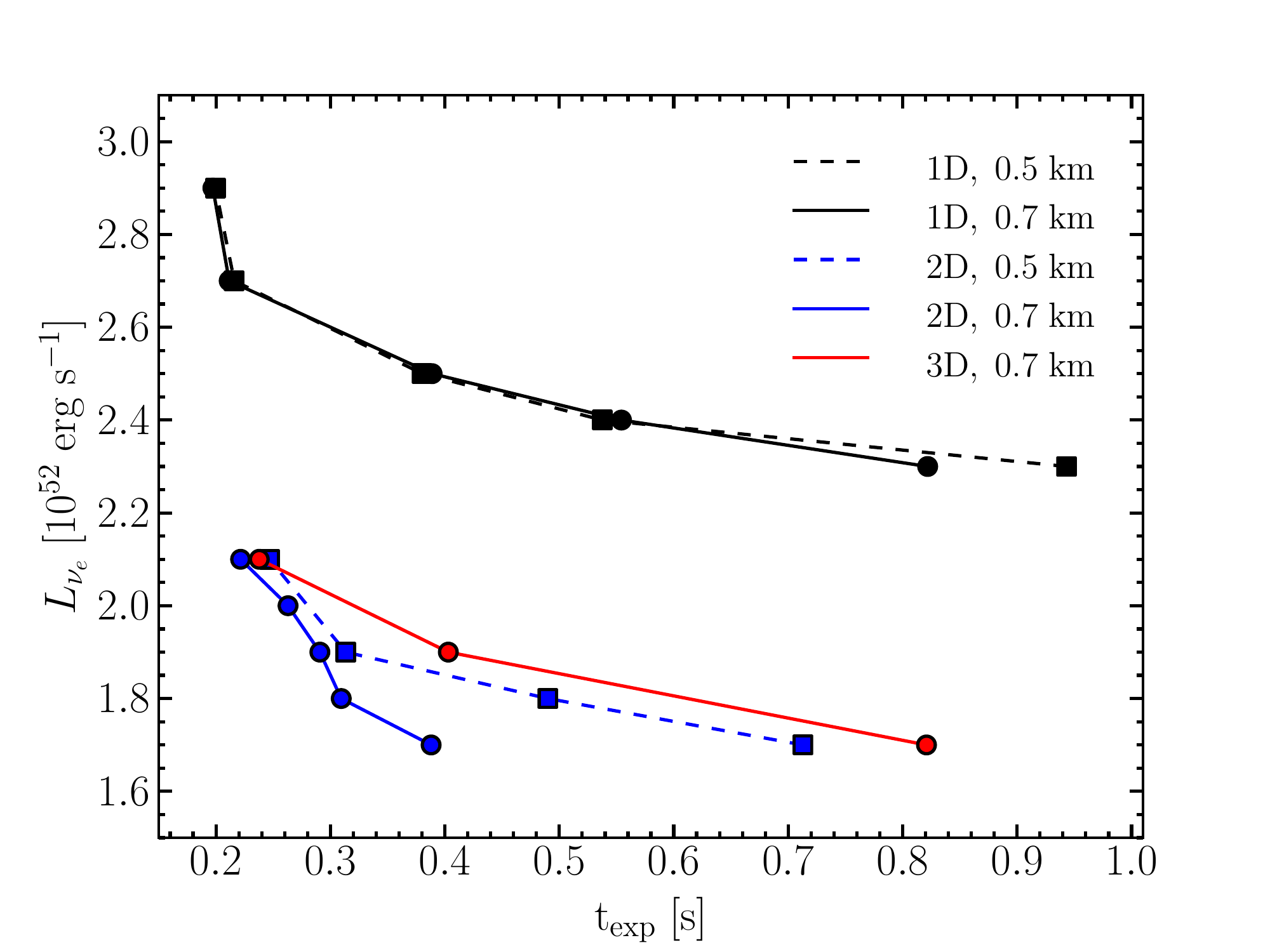}
\end{tabular}
\caption{Critical neutrino luminosity curves in both mass accretion
  rate (left) and post-bounce time (right) for our set of 1D, 2D, and
  3D simulations with the 15 $M_\sun$ progenitor.  We reproduce the
  common result that the critical curves are lower in 2D than in 1D,
  however, we find that for our fiducial resolution of 0.7 km the 3D
  critical curves are {\it higher} than the 2D curves.  Thus, our
  results indicate that obtaining explosion is more difficult in 3D.
  Critical curves for increased resolution 1D and 2D simulations are
  also shown.  Increasing resolution in 1D results in almost no
  difference in the curves, whereas in 2D the higher-resolution
  simulations explode later and the curves are much closer to the 3D
  counterparts.}
\label{fig:mdot}
\end{figure*}

We have run a series of 1D, 2D, and 3D simulations in which we varied
the driving neutrino luminosity.  We start in the pre-collapsed
progenitor phase and follow the evolution through collapse, bounce,
shock stagnation and eventual revival. Figure \ref{fig:volRender}
shows a volume rendering of entropy and the shock surface in a 3D
simulation at 850 ms post-bounce.  In Table \ref{table:results} we
give the explosion delay times for our series of simulations and
Figure \ref{fig:rshock} shows the average shock radii as a function of
time post-bounce for a number of our simulations.  Figure
\ref{fig:mdot} shows the critical luminosity curves as functions of
both post-bounce explosion time and mass accretion rate at explosion.
We consider a model to have exploded once the average shock radius
exceeds 400 km and does not subsequently fall back below this value
\citep[as in][]{Nordhaus:2010ct,Hanke:2012dx}, though other metrics,
such as reaching a critical value of the ratio of advection time to
heating time in the gain region \citep[e.g.,][]{Fernandez:2012kg} or
satisfying the `ante-sonic' condition \citep{Pejcha:2012cw} may be
used \citep[for a comparison of the difference between these metrics,
see][]{Dolence:2013iw}.  We find that the critical luminosity curve is
lowered in multidimensional simulations as compared with
spherically-symmetric simulations, consistent with all previous
similar studies \citep{Murphy:2008ij, Nordhaus:2010ct, Hanke:2012dx,
  Couch:2013df}.  When comparing 2D to 3D, however, we find
interesting and heretofore unprecedented behavior: at our fiducial
resolution the 2D simulations consistently explode {\it earlier} than
3D simulations at the same neutrino luminosity.  Figure
\ref{fig:rshock} shows that for a given neutrino luminosity the
average shock radius expands more quickly in 2D than 3D.
 
In 2D, the explosion time for a given luminosity is sensitive to the
grid resolution used.  Increasing the finest grid resolution to 0.5 km
and reducing the radial refinement limiter, $\eta$, to 0.94\% results
in a 2D criticality curve much more similar to the 3D curve at the
fiducial resolution of 0.7 km (and $\eta=1.25$\%).  Increasing the
resolution in 1D simulations results in almost {\it no change} in the
explosion times as a function of neutrino luminosity.  This very
importantly indicates that the cause of the resolution dependence is
connected to an intrinsically multidimensional process.  At present,
we lack the necessary computational resources to carry out a
resolution study in 3D, though our results certainly indicate the
necessity of such a study.  \citet{Hanke:2012dx} do carry out a
resolution study including 3D and also find that the explosion times
are very dependent on grid spacing.  For simulations with 400
unequally-spaced radial zones, Hanke et al. find that increasing the
angular resolution of their spherical grid results in earlier
explosions in 2D but later explosions in 3D.  Considering simulations
with at least 600 radial zones, the dependence of the explosion times
on resolution is not always consistent amongst their results;
increasing the angular resolution delays explosion in 2D for some of
their models.  It is important to note that their fiducial resolution
($3\fdg$) is substantial coarser than ours ($\sim 0\fdg54$) and the
finest resolution they use in 3D ($1\fdg5$) is still coarser than our
resolution.

\section{Interpretation and Analysis}
\label{sec:interpret}
\subsection{Shock Expansion Driven by Buoyancy}

\begin{figure}
\centering
\begin{tabular}{cc}
\includegraphics[width=1.75in,trim = 4in 1.75in 4in 0in, clip]{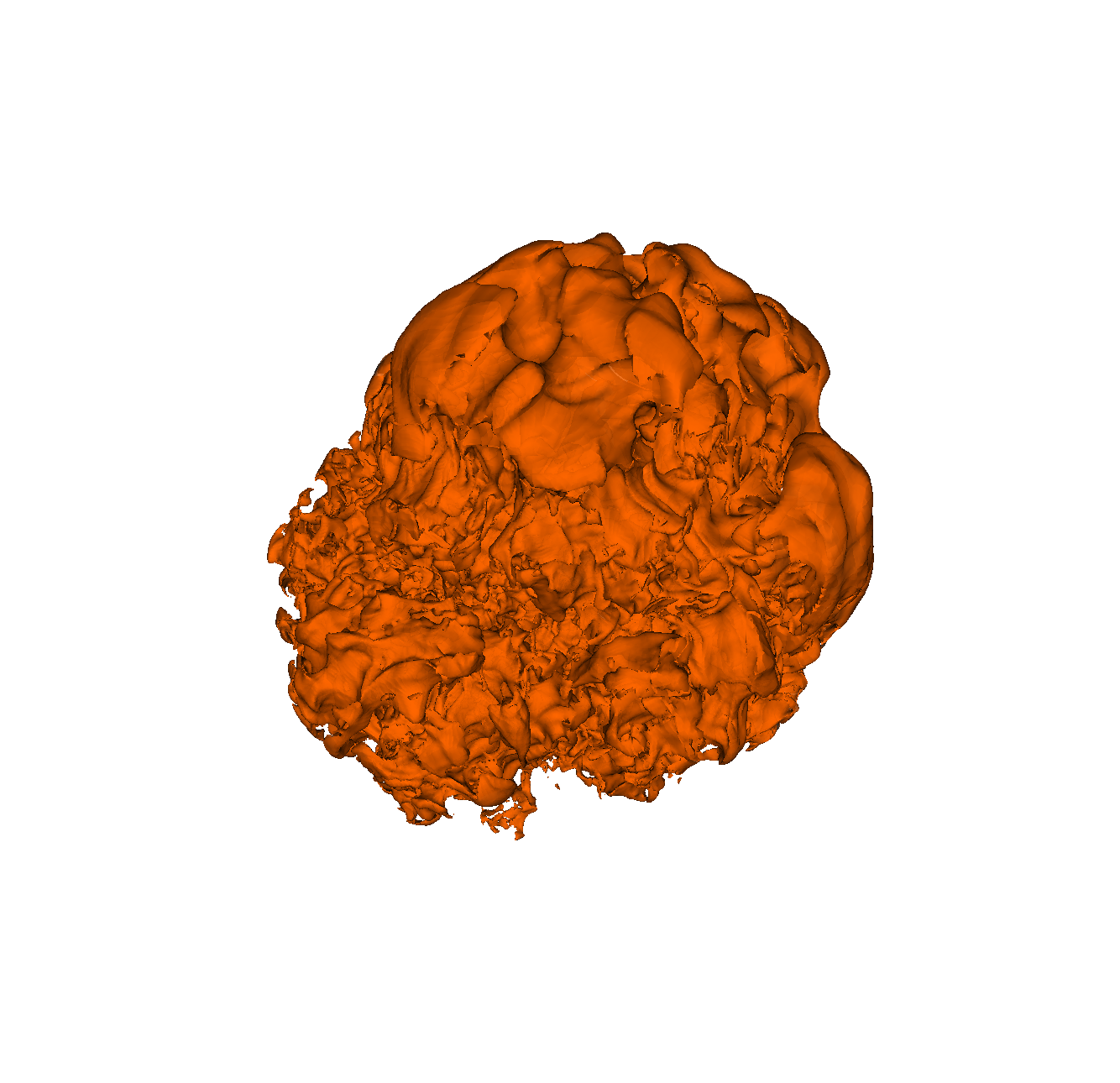}
\includegraphics[width=1.75in,trim = 4in 1.75in 4in 0in, clip]{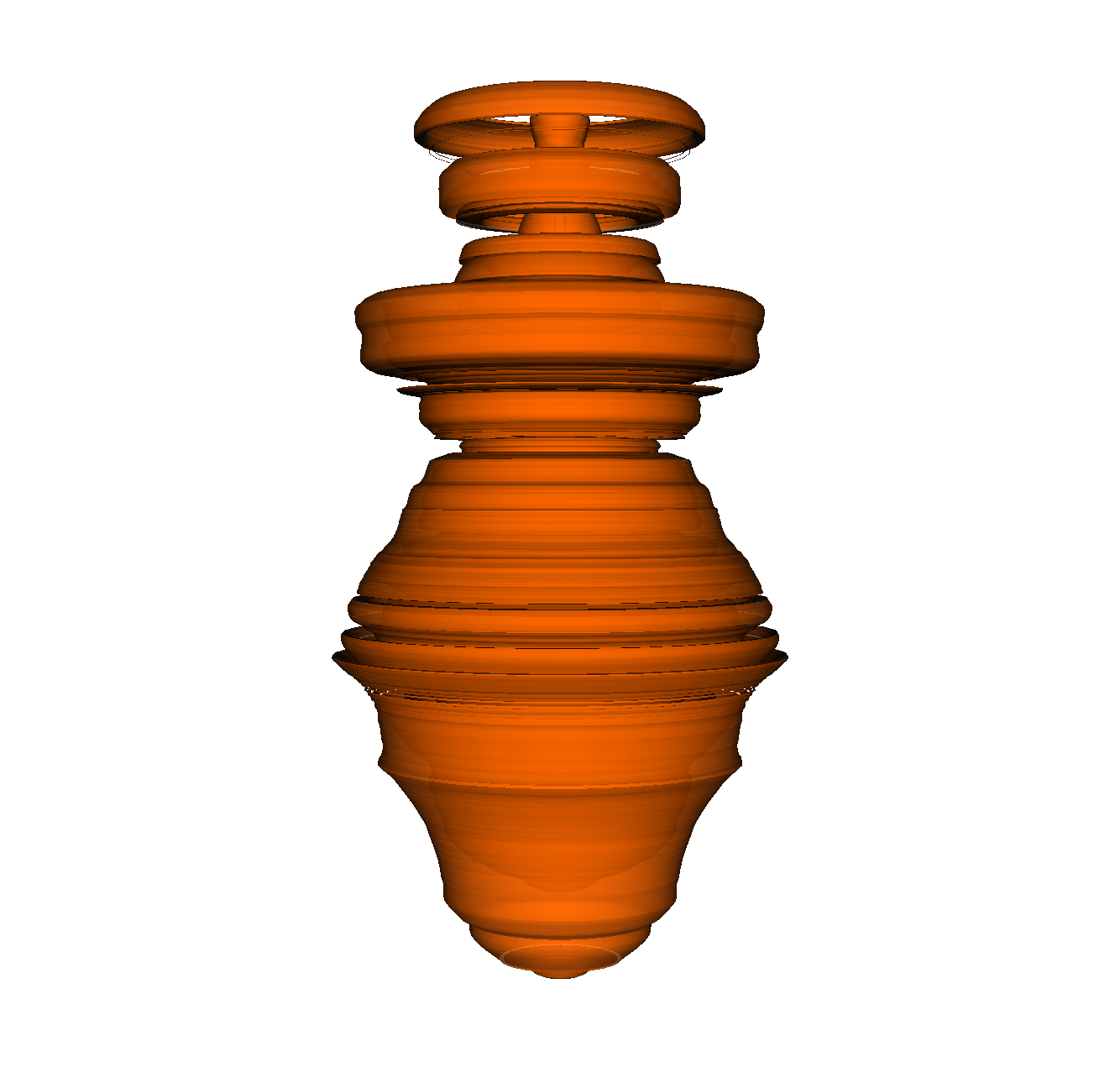}
\end{tabular}
\caption{Constant entropy contours for a value of 14 $k_B$
  baryon$^{-1}$ for models with $L_{\nu_e,52}=1.7$ at the respective
  times of explosion (see Table \ref{table:results}).  The left panel
  shows the 3D data and the right shows the 2D data revolved about the
  symmetry axis.  As discussed in the text, 2D simulations show
  buoyant plumes that have much smaller surface area-to-volume ratios
  than for 3D.}
\label{fig:contours}
\end{figure}

So then, what is the explanation of our results?  That is, why is it
that our 2D simulations explode earlier than our 3D? and why does
increasing the resolution in 2D result in later explosions?  As we
will argue in the following sections, our results indicate that
neutrino-driven buoyant convection is the dominant instability that
encourages shock expansion for this progenitor, particularly in 3D.
Similar arguments have been made recently by \citet{Burrows:2012gc}
and \citet{Murphy:2013eg}.  In this picture, accreting gas in the gain
region absorbs neutrino energy eventually becoming buoyant and rises
toward the shock where this buoyant energy is used to push the shock
further out.  The speed at which a plume rises will be determined by
the competition between the plume's buoyancy, determined by the amount
of neutrino energy it has absorbed, and the drag force from cold gas
traveling downward through the gain region \citep[see,
e.g.,][]{{Thompson:2000gd}, Dolence:2013iw}.  A plume's buoyancy will
be proportional to its volume since the neutrino energy absorption
rate will scale as the solid angle of the plume times its neutrino
optical depth.  The drag force pushing back against the buoyant plume
will scale as the surface area of the plume.  Therefore, smaller
plumes with greater surface area-to-volume ratios will rise more
slowly than larger plumes.

\begin{figure*}
\centering
\begin{tabular}{cc}
\includegraphics[width=3.5in,trim= 3.5in 1.75in 3.5in .75in,clip]{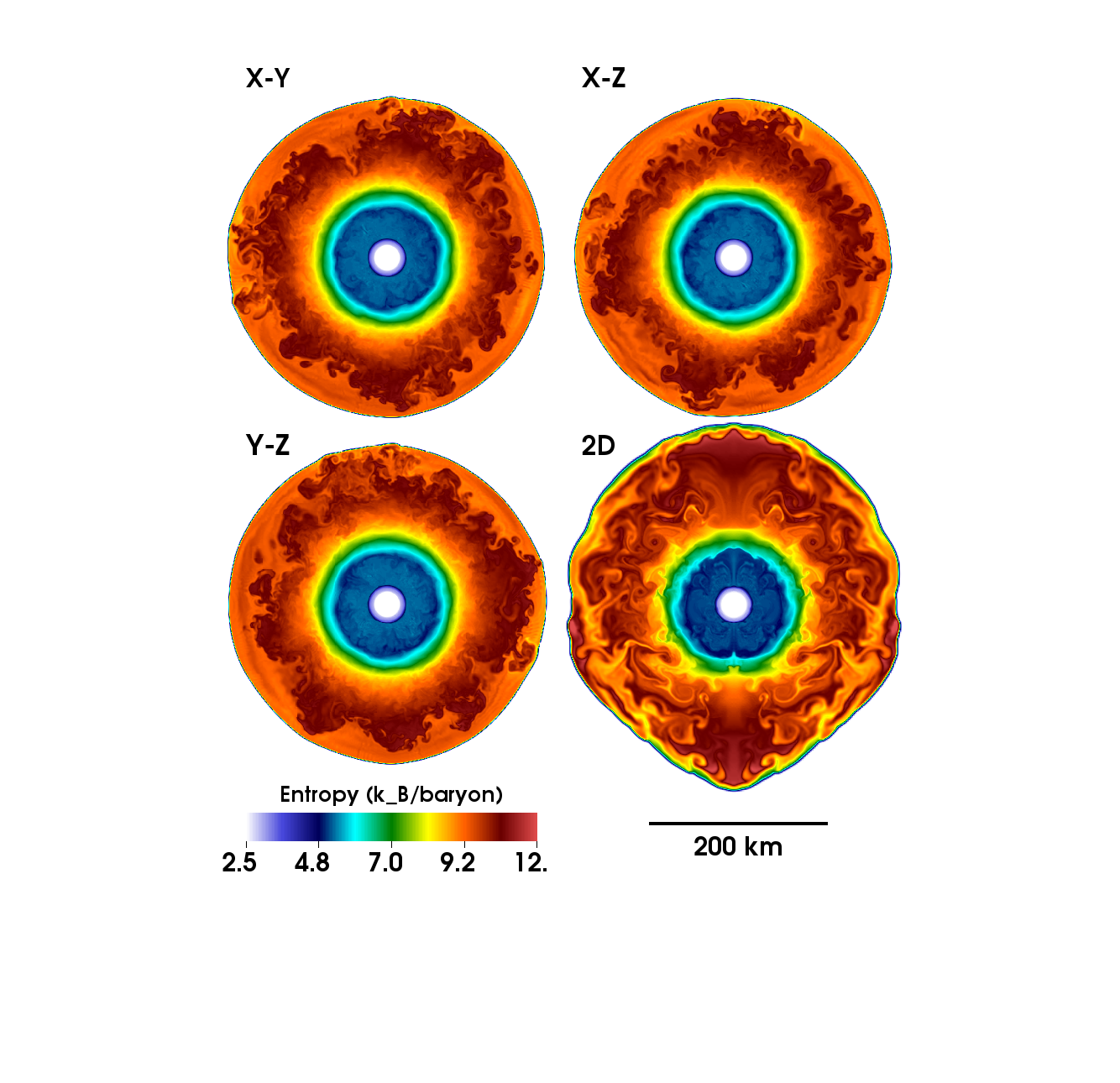}
\includegraphics[width=3.5in,trim= 3.65in 1.8in 3.8in 0.8in,clip]{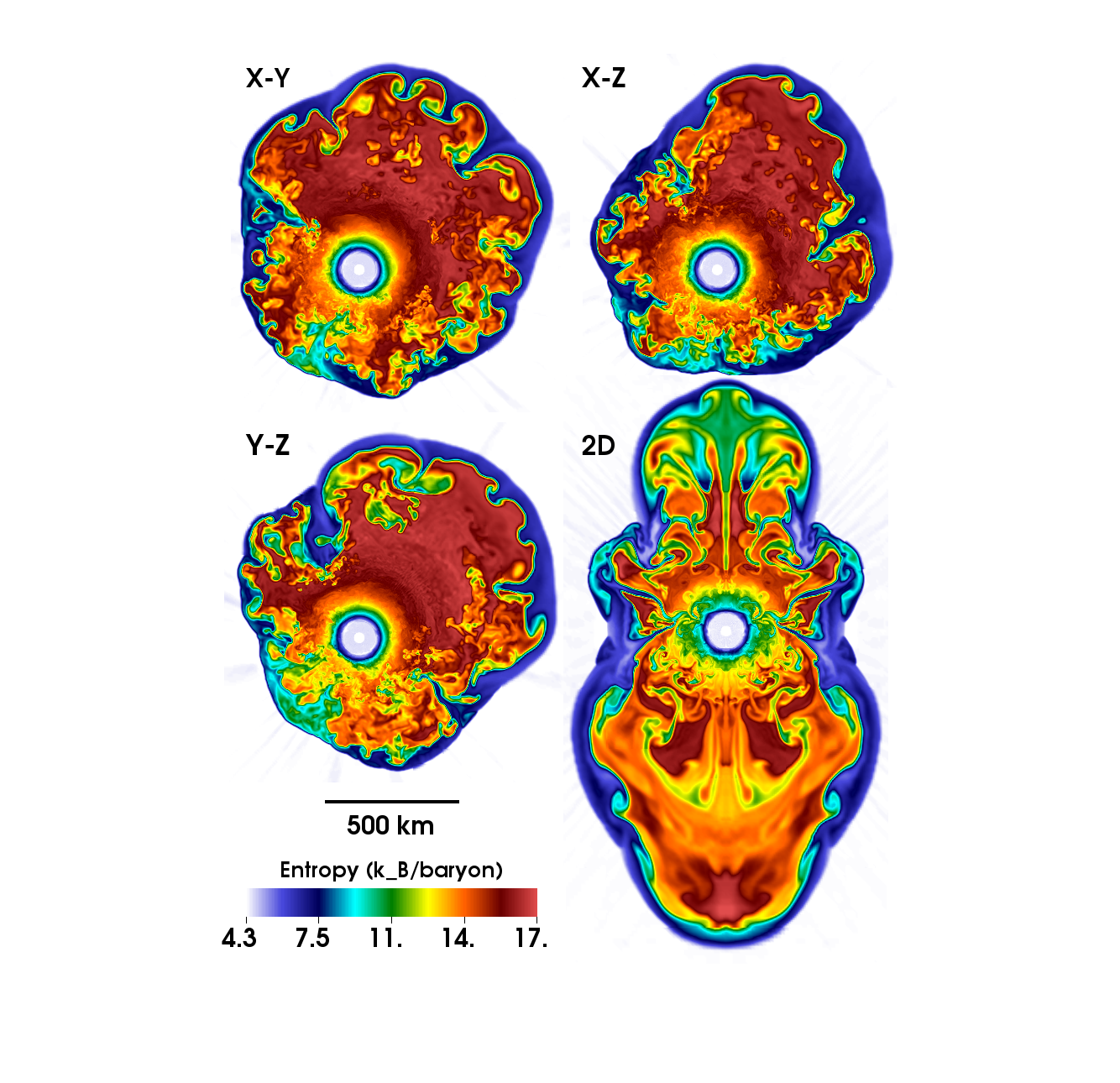}
\end{tabular}
\caption{Entropy pseudo color plots for 3D and 2D simulations with
  $L_{\nu_e,52}=1.7$ at 100 ms post-bounce (left) and time of explosion
  (right).  For 3D, three orthogonal slice planes are shown. By 100 ms
  the shock structure in 3D is still very spherical and high-entropy
  buoyant plumes are just starting to reach the shock.  In 2D at 100
  ms, elongation along the symmetry axis is already evident,
  particularly in the southern hemisphere, and the convective
  structures are larger and more coherent.  At explosion time, 388 ms
  and 821 ms in 2D and 3D, respectively, the character of the shock
  and the buoyant convection behind it is completely divergent between
  2D and 3D.  The 2D explosion is characteristically dipolar and
  dominated by a few, large buoyant plumes (or arguably $\ell=1$
  SASI).  In 3D, the shock does not show a dominant low-order shape
  and the convective plumes show much more small scale structure.
  There are also a greater number of low-entropy down flows in 3D.
  These fundamental differences between 2D and 3D result in convective
  plumes that have much greater surface area-to-volume ratios in 3D.
  This results in a greater drag-to-buoyant force ratio which, for
  buoyancy-dominated shock expansion, leads to slower shock expansion
  in 3D.}
\label{fig:3Dv2Dslices}
\end{figure*}

To see this, consider a spherical buoyant bubble of radius $r_b$ at a
distance from the coordinate origin $R_b$.  The instantaneous buoyant
force on this bubble is provided by neutrino radiation and so must be
equal to the neutrino radiation force: $F_\nu \sim \tilde{\sigma_\nu}
\rho_b r_b^3 L_\nu / c R_b^2$, where $\tilde{\sigma_\nu}$ is an
effective neutrino cross-section that includes any geometric
constants, $\rho_b$ is the density of the bubble, and $c$ is the speed
of light.  The instantaneous drag force on the bubble is $F_d \sim
\tilde{C_d} \rho_b v^3 r_b^2$, where $\tilde{C_d}$ is a drag
coefficient that contains any geometric constants, and $v$ is the
bubble's velocity relative to the background accretion flow.  The
ratio of the buoyant force to the drag force on the bubble is then
\beq \frac{F_\nu}{F_d} \sim \frac{\tilde{\sigma_\nu} L_\nu
  r_b}{\tilde{C_d} v^2 c R_b^2}.
\label{eq:ratio}
\eeq 
Thus, this ratio increases with bubble size, $r_b$, and larger
bubbles or plumes will rise faster than smaller.

In 2D, the typical plume size is larger than in 3D for multiple
reasons.  First, the axial symmetry intrinsic to 2D cylindrical
coordinates means that off-axis plumes are really rings.  In 3D, such
plume rings are unstable and will break up into many {\it smaller}
plumes.  In Figure \ref{fig:contours} we demonstrate this difference
in plume scale with constant entropy contours in 3D (left) and 2D
revolved around the symmetry axis (right).  In 2D the forced symmetry
results in very large ``3D'' plumes whereas in 3D no such large scale
plumes can exist.  These contours are plotted at the respective
explosion times in correspondence with the right panel of Figure
\ref{fig:3Dv2Dslices}. Second, the symmetry axis in 2D encourages the
growth of large plumes along it, due either to the action of low-order
modes of the SASI or that of buoyant convection.  In Section
\ref{sec:shock} we show that the amplitudes of low-order spherical
harmonic modes of the shock deformation are reduced in 3D as compared
to the 2D case.  Third, the ``inverse'' turbulent energy cascade in 2D
\citep{Kraichnan:1967jk} will pump energy to larger scales whereas in
3D the ``forward'' energy cascade will send energy to smaller scales
(see Section \ref{sec:turb}).

\begin{figure*}
\centering
\includegraphics[width=7in,trim = .25in 0.1in .25in 0in,clip]{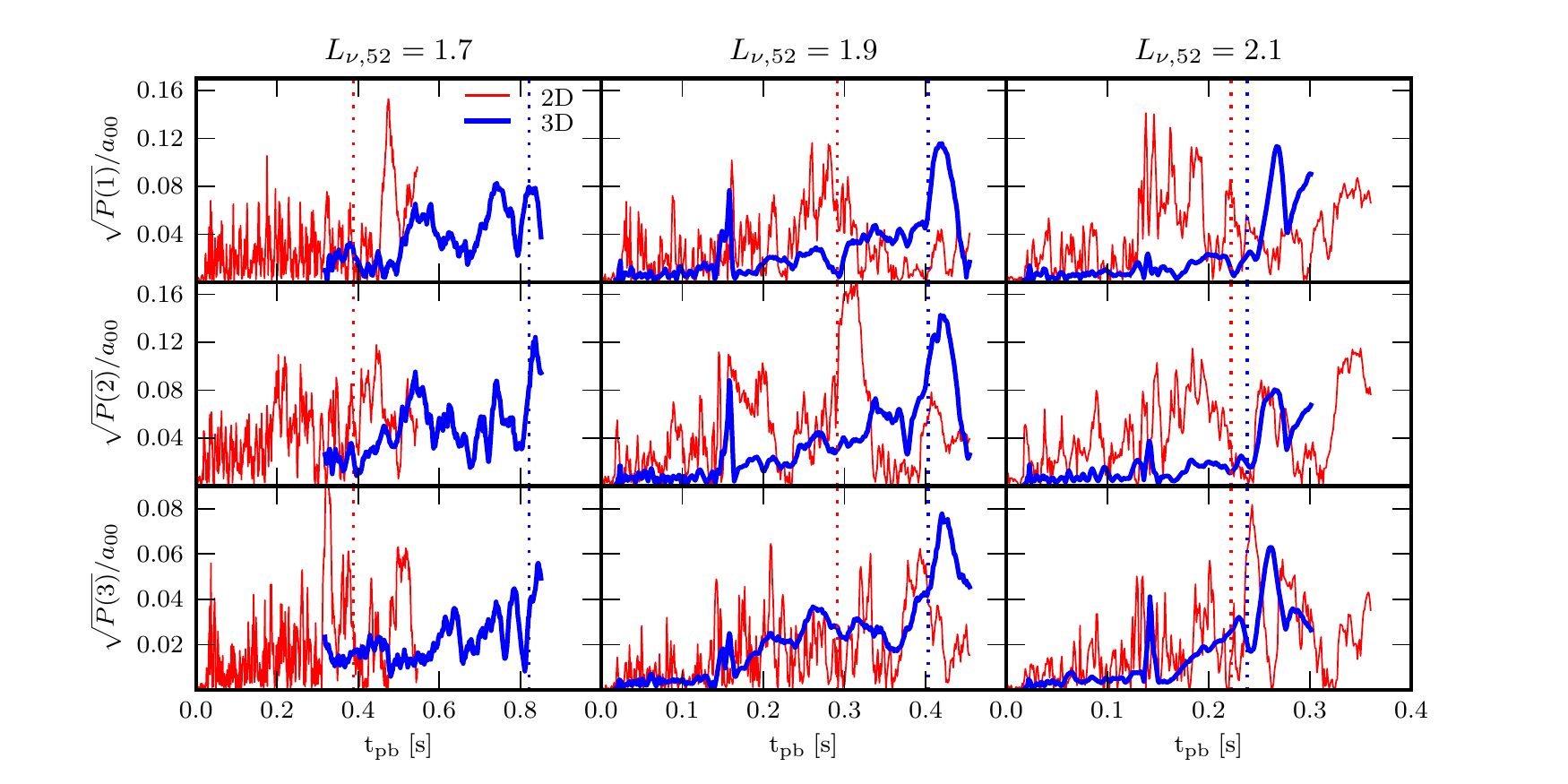}
\caption{Normalized power in the first three spherical harmonic modes
  of the shock deformation in 2D (red) and 3D (blue) for three
  neutrino luminosities.  The ``power" is the squared spherical
  harmonic coefficient summed over all $m$'s, so in 2D where the $m$'s
  cancel $P(\ell) = a^2_{\ell 0}$.  The amplitudes for 3D are
  generally smaller than for 2D.  The data necessary to compute the
  shock spherical harmonics for the 3D model with $L_{\nu_e,52}=1.7$
  prior to $t=0.3$ s was, mistakenly, not saved.}
\label{fig:spharm}
\end{figure*}

In Figure \ref{fig:3Dv2Dslices} we compare entropy slices between 3D
and 2D simulations with $L_{\nu_e,52}=1.7$.  The left half of Figure
\ref{fig:3Dv2Dslices} is at a time of 100 ms post-bounce and the right
half is at the time of explosion, 821 ms for 3D and 388 ms for 2D.  At
100 ms, the 3D simulation shows a shock that is still nearly spherical
and developing convection behind it.  The largest convective plumes
are just reaching the shock perturbing it's spherical structure and
stochastically pushing it out in radius.  The 2D simulation is
similar, but the developing convection in the gain region is visibly
more coherent and vortex-like.  Again, in 2D these vortical convective
cells truly represent three-dimensional rings.  Also in 2D, the
influence of the symmetry axis is already apparent as the shock is
becoming elongated along it, particularly along the southern axis.  At
the time of explosion, the 2D and 3D structures have diverged
significantly.  The 2D explosion shows the characteristic ``bipolar''
shock structure and is dominated by about three large buoyant plumes,
one each along the axes and one, somewhat smaller buoyant plume north
of the equator.  The 3D structure is quite different: many smaller
buoyant plumes exist and are more evenly distributed in solid angle
and the shock, while certainly not perfectly spherical, does not show
large-amplitude low-order deformation as in the 2D case.  The salient
point is that Figures \ref{fig:contours} and \ref{fig:3Dv2Dslices}
clearly demonstrate that the surface area-to-volume ratio of the
buoyant plumes is much higher in 3D than in 2D.  Thus, due to the
greater amount of drag relative to buoyant force the plumes will rise
more slowly in 3D than in 2D and will, therefore, encourage a slower
growth of the average shock radius and later explosion times.

\subsection{Character of the Shock Motion}
\label{sec:shock}

Multidimensional phenomena such as neutrino-driven convection and the
SASI naturally result in aspherical shock motion in CCSNe.  Linear
analysis indicates that the most unstable modes of the SASI will be
low-order, $\ell \approx$ 1 \citep{Foglizzo:2007cq,Guilet:2012ib}.
Three-dimensional simulations \citep{Iwakami:2008dw, Iwakami:2009er}
and 2D $R-\phi$ axisymmetric simulations \citep{{Blondin:2007bf}}
tailored to study the SASI show that the SASI develops an $m=1$
``spiral'' mode that can be considered the superposition of two or
more out-of-phase $\ell=1$ modes
\citep{Fernandez:2010ko}. Neutrino-driven convection will excite much
higher-$\ell$ modes of the shock motion and we have in the previous
sections argued that such buoyant convection is the dominant
instability that encourages expansion of the shock in the present
simulations.  In this section we justify this assertion by analyzing
the shock motion via spherical harmonic decomposition.  Our approach
is comparable to that of \citet{Burrows:2012gc}.  A very thorough
study of the character of the shock motion in 3D is also presented by
\citet{Ott:2013gz}.  Their 3D GR simulations with neutrino leakage are
the most physically-detailed for which the shock spherical harmonics
have been investigated.  They find that high-order neutrino-driven
convection dominates the low-order development of the SASI, which is
evident but weak in their calculations.

We can decompose some scalar quantity, $X$, into spherical harmonic
components with coefficients 
\beq a_{\ell m} = \oint X(\theta,\phi)
Y_\ell^m(\theta,\phi) d\Omega,
\label{eq:alm}
\eeq
where the spherical harmonics are
\beq
Y_\ell^m = 
\begin{cases} 
\sqrt{2} N_\ell^m P_\ell^m(\cos\theta) \cos m\phi &		m>0,\\
N_\ell^0 P_\ell^0(\cos \theta) &				m=0,\\
\sqrt{2} N_\ell^{|m|} P_\ell^{|m|}(\cos\theta) \sin |m|\phi &	m<0,
\end{cases}
\eeq 
and 
\beq N_\ell^m = \sqrt{\frac{2\ell + 1}{4\pi}
  \frac{(\ell-m)!}{(\ell+m)!}}.  
\eeq 
The spherical harmonics are
computed from the associated Legendre polynomials, $P_\ell^m$, and we
use the physics convention for spherical coordinates: $\theta$ is the
polar angle and $\phi$ is the azimuthal angle.

In order to study the shock behavior, we decompose the shock surface
$R_s(\theta,\phi)$ according to equation (\ref{eq:alm}) and define a
spherical harmonic ``power'': \beq P(\ell) = \sum_{m=-\ell}^{\ell}
a_{\ell m}^2, \eeq where we follow \citet{Burrows:2012gc} in
normalizing the coefficients by a factor
$(-1)^{|m|}/\sqrt{4\pi(2\ell+1)}$ such that $a_{00} = \langle R_s
\rangle$, $a_{11} = \langle x_s \rangle$, $a_{1-1} = \langle y_s
\rangle$, and $a_{10} = \langle z_s \rangle $. In 2D, all $m \neq 0$
components cancel so that $P(\ell) = a_\ell^2$.  We use an accurate
shock finding algorithm in FLASH to track and flag zones that are
within the shock \citep{Balsara:1999ee}.

We show in Figure \ref{fig:spharm} the first three spherical harmonic
powers of the shock surface, normalized by the average shock radius,
for both 2D and 3D simulations at various neutrino luminosities as
functions of time.  We find that the amplitudes of the low-order
harmonics are significantly reduced in 3D as compared to 2D,
particularly prior to the onset of explosion (marked by vertical
dotted lines in Figure \ref{fig:spharm}).  In 3D, the amplitudes are
especially small prior to accretion of the Fe/Si interface at around
125 ms.  When this occurs, the higher entropy in the Si shell
encourages rapid shock expansion (Fig. \ref{fig:rshock}) that excites
substantial shock deformation.  This transient high-amplitude spike
fades away, however, after the entirety of the Fe/Si interface is
accreted through the shock.  We find a general trend that the later
the explosion time, the higher the $P(1,2,3)$ at explosion.  For
example, the 3D model with $L_{\nu_e,52}=2.1$ explodes with very small
$P(1,2,3)$ and only after explosion sets in do the amplitudes grow to
larger values.

Of note, and not evident in Figure \ref{fig:spharm}, is that between
50 and 100 ms we find a low-amplitude spiral mode of the shock
deformation.  This spiral mode is damped once neutrino-heated plumes
reach and impinge upon the shock at around 100 ms (see
Fig. \ref{fig:3Dv2Dslices}).  Also, we have computed the spherical
harmonic powers of the shock deformation for the higher-resolution 2D
simulations and find no significant difference from the fiducial
resolution 2D simulations.

\subsection{Explosion Indicators}
\label{sec:indicators}

\begin{figure}
\centering
\includegraphics[width=3.45in,trim= 0.05in 0in 0.4in 0.25in, clip]{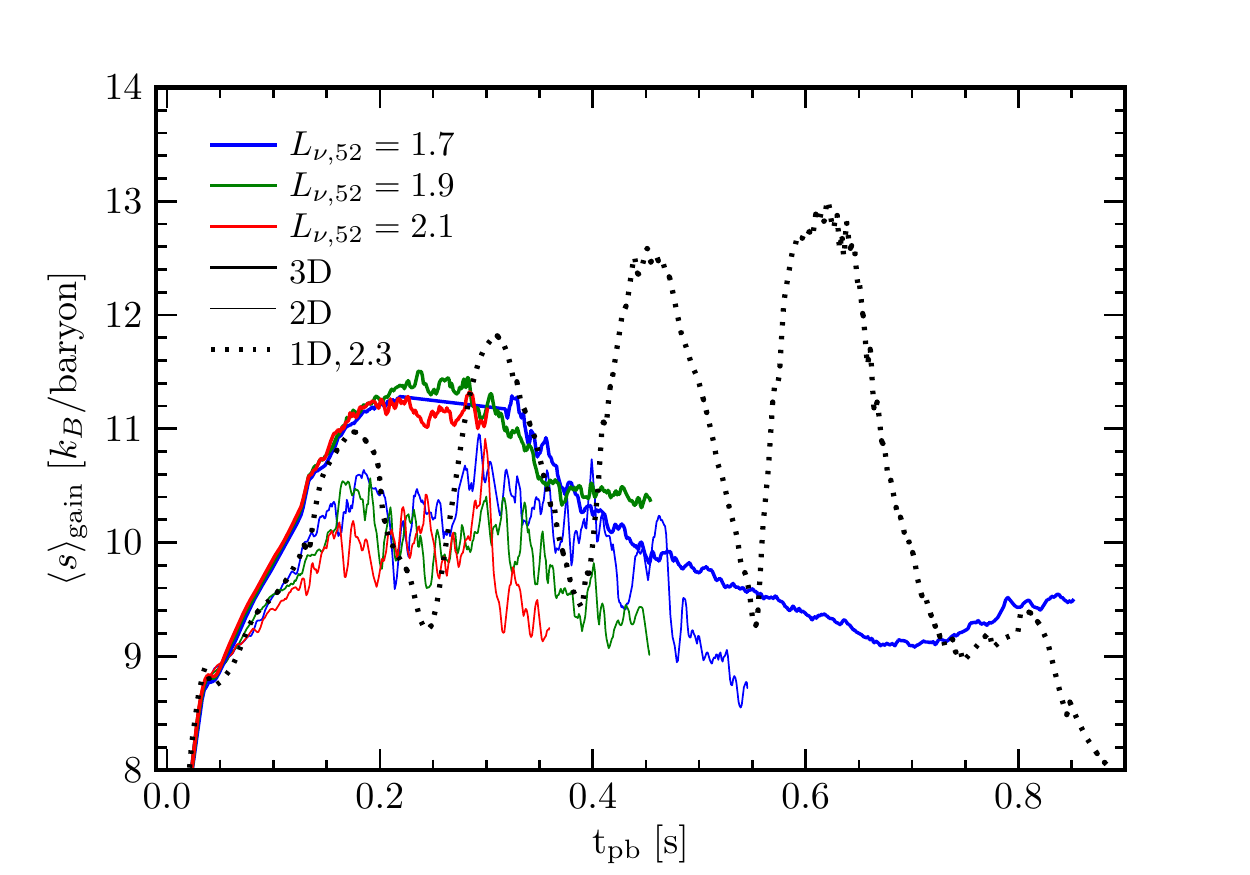}
\caption{Mass-averaged entropy in the gain region.  The average
  entropies in 3D are systematically higher than in 2D, a result of
  the smaller mass in the gain region for 3D.  We do not find that
  average gain region entropy is an indicator of proximity to
  explosion. Shown for comparison are the 1D data with
  $L_{\nu_e,52}=2.3$.}
\label{fig:entropy}
\end{figure}

\begin{figure}
\centering
\begin{tabular}{cc}
\includegraphics[width=3.45in,trim= 0.05in 0in 0.4in 0.25in, clip]{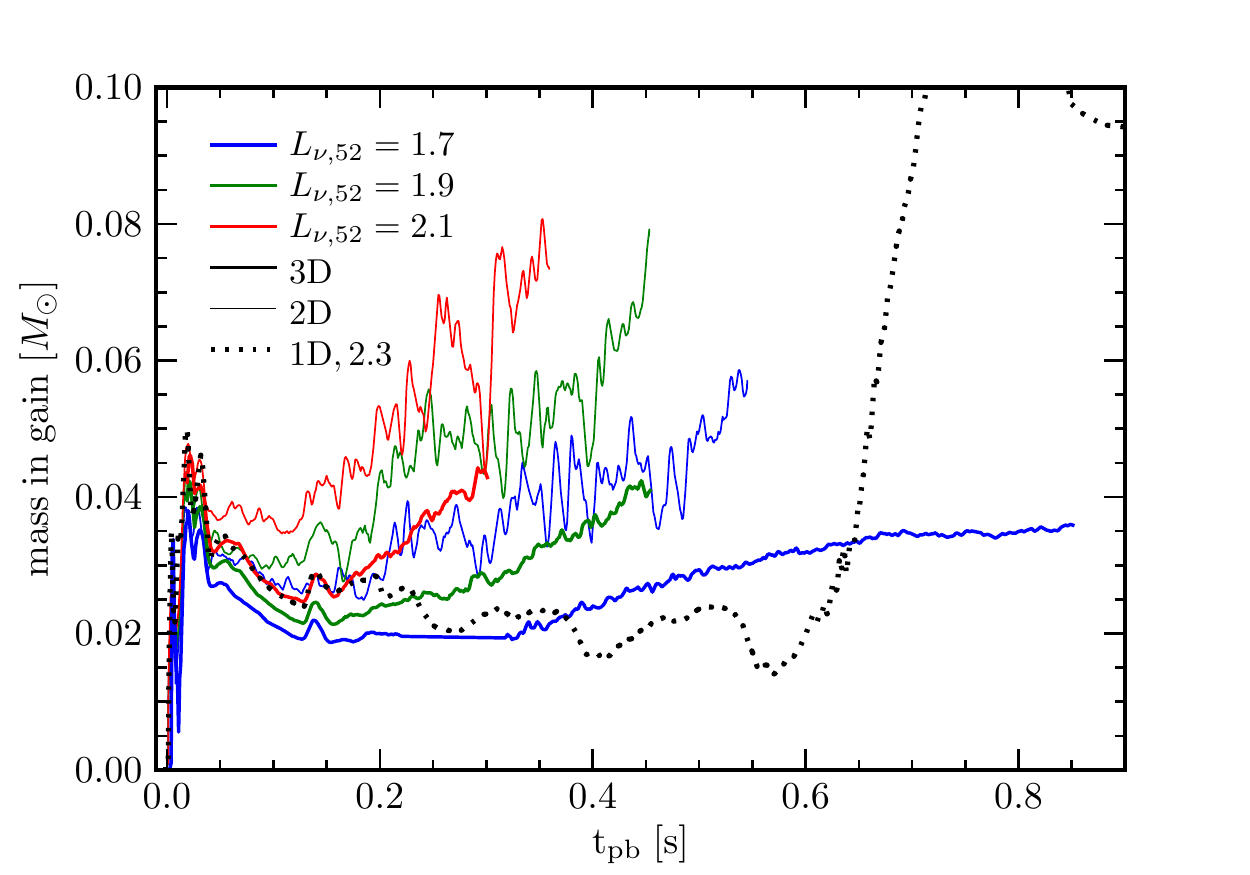}
\end{tabular}
\caption{Mass in the gain region as a function of time.  Gain region
  mass is higher in 2D than in 3D, resulting in a higher net heating
  rate (Fig. \ref{fig:heating}) and a lower mass-averaged entropy
  (Fig. \ref{fig:entropy}). }
\label{fig:mass}
\end{figure}

\begin{figure}
\centering
\begin{tabular}{cc}
\includegraphics[width=3.45in,trim= 0.05in 0in 0.4in 0.25in, clip]{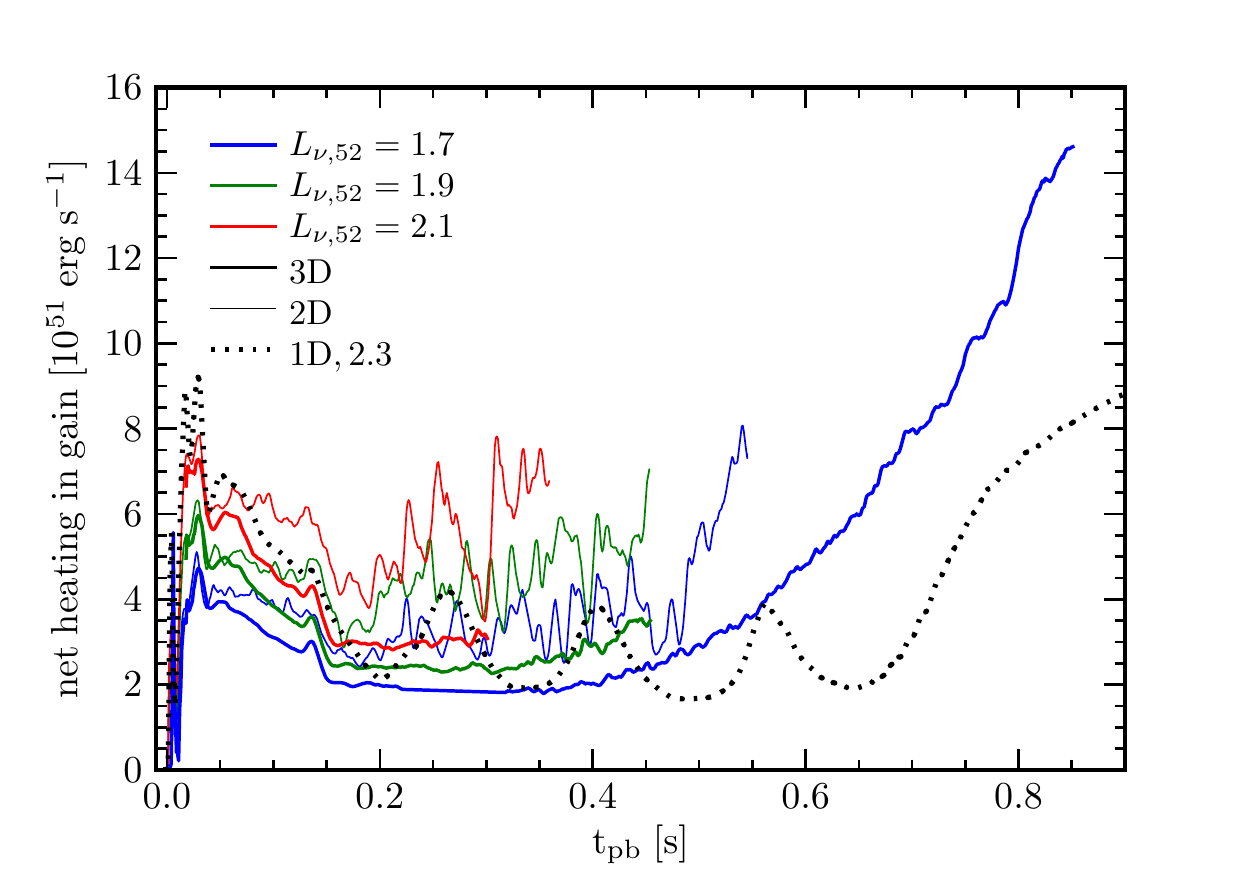}
\end{tabular}
\caption{Net neutrino heating rate in the gain region.  Because of the
  greater gain region mass in 2D, the net heating is also higher.  The
  net heating rises dramatically at late times (thick blue line)
  because the average temperature in the gain region drops, reducing
  the neutrino cooling rate (eq. \ref{eq:cool}). }
\label{fig:heating}
\end{figure}

A number of quantities have been identified as possible indicators or
predictors of explosion in core-collapse supernovae.
\citet{Nordhaus:2010ct} argued that a higher average entropy in the
gain region correlates with likelihood of explosion and found a clear
dimensional hierarchy in this quantity: 3D $>$ 2D $>$ 1D.
\citet{Hanke:2012dx}, however, suggest that there exists only a small,
possibly insignificant, dependence of the gain region average entropy
on dimensionality.  In Figure \ref{fig:entropy} we plot the
mass-averaged entropy in the gain region for our 2D and 3D
simulations.  We define the gain region as any part of the simulation
domain that has a net positive neutrino heating as determined by the
difference between equations \ref{eq:heat} and \ref{eq:cool}.  We find
that there is a clear hierarchy between 2D and 3D with 3D having
significantly higher average entropies than 2D.  The 1D average
entropy, however, does not fit the dimensional hierarchy.  Our results
point out that average entropy in the gain region is {\it not} a good
indicator of explosion likelihood as we find that 2D explodes before
3D yet still has smaller average entropies.  The average entropy in
the gain region is rather a measure of the time-integrated specific
energy absorbed.  Thus, the lower average entropies in 2D reflect that
the gain region in 2D contains a greater amount of mass than in 3D, as
shown in Figure \ref{fig:mass}.  This higher gain region mass is also
reflected in a higher integrated heating rate
(Fig. \ref{fig:heating}).  A higher net heating rate should,
intuitively, result in a greater rate of shock expansion.

\begin{figure}
\centering
\begin{tabular}{cc}
\includegraphics[width=3.45in,trim= 0.05in 0in 0.4in 0.25in, clip]{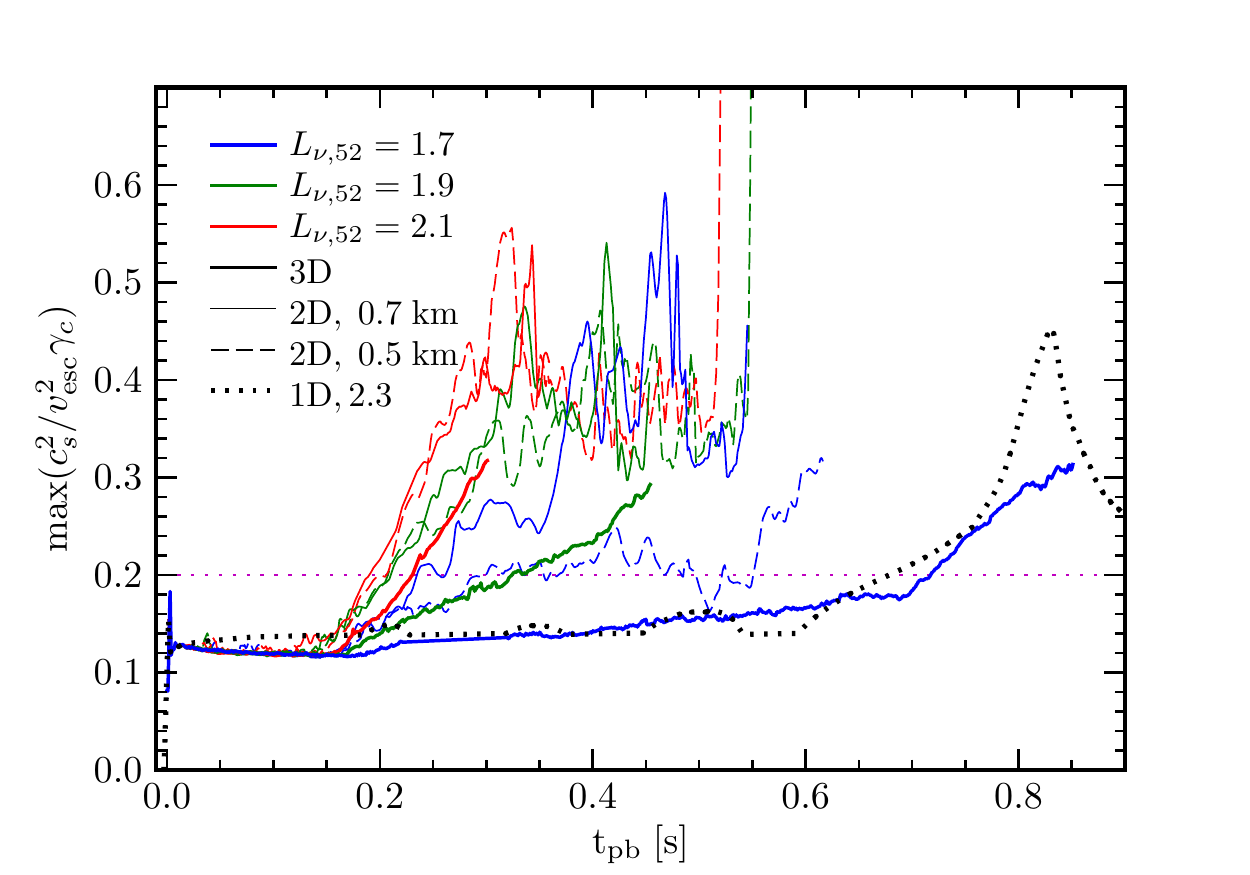}
\end{tabular}
\caption{Maximum value of the ratio of the squared adiabatic sound
  speed to the squared escape velocity times adiabatic index, the
  `antesonic' condition.  We find that the antesonic condition is a
  good predictor of proximity to explosion, although our data indicate
  that the critical value of $\sim 0.2$ suggested by
  \citet{Pejcha:2012cw} is somewhat low.  We also plot the data for
  the high-resolution 2D simulations.  The antesonic condition is a
  better predictor of the delayed explosion time for the
  high-resolution simulations than any of the other integral
  quantities plotted in Figs. \ref{fig:entropy} --
  \ref{fig:heating}. We also show the critical ratio for the 1D model
  with $L_{\nu_e,52}=2.3$ (black dotted line).  The 1D data look very
  similar to the 3D data for $L_{\nu_e,52}=1.7$ and peak at explosion
  time.}
\label{fig:antes}
\end{figure}

\citet{Pejcha:2012cw} suggest that instability to runaway shock
expansion sets in when the maximum value of the sound speed squared to
the escape velocity squared reaches a critical value, the so-called
`antesonic' condition.  For an isothermal equation of state, the
antesonic condition is $\max(c_s^2/v_{\rm esc}^2) = 0.19$, where $c_s$
is the adiabatic sound speed and $v_{\rm esc}$ is the escape speed.
For a polytropic equation of state, the antesonic condition becomes
$\max(c_s^2/v_{\rm esc}^2) = 0.19\Gamma$, where $\Gamma$ is the
polytropic index.  For the completely general EOS we use, we modify
the polytropic antesonic condition to be \beq \max(c_s^2/v_{\rm esc}^2
\gamma_c) = 0.19,
\label{eq:antes}
\eeq where $\gamma_c$ is now the varying adiabatic index given by the
EOS.  In Figure \ref{fig:antes} we plot the left hand side of equation
(\ref{eq:antes}) as a function of time for our simulations.  We find
that the antesonic condition is a good indicator of the beginning of
accelerated shock expansion, although a critical value of 0.2 may be a
little low \citep[see also][]{Hanke:2012dx,Dolence:2013iw}.  We see
that an antesonic value of about 0.3 corresponds very well with our
definition of the explosion time, when the average shock radius
exceeds 400 km.  Also, the antesonic value for our 2D simulations
exceeds the critical value before the 3D simulations, echoing our
result that 2D explodes before 3D.

In Figure \ref{fig:antes} we also plot the antesonic value for 2D
simulations with higher resolution.  Interestingly, of the quantities
shown in Figures \ref{fig:entropy} - \ref{fig:antes} the antesonic
condition shows the most noticeable differences between the 0.7 km and
0.5 km resolution 2D simulations.  The difference is most notable for
the simulations with $L_{\nu_e,52}=1.7$ where we see that based on the
antesonic condition we would expect the higher resolution simulation
to explode later as, in fact, it does.

\subsection{Non-Radial Motion and Turbulence}
\label{sec:turb}

\begin{figure}
\centering
\includegraphics[width=3.45in,trim= 0.05in 0in 0.4in 0.25in, clip]{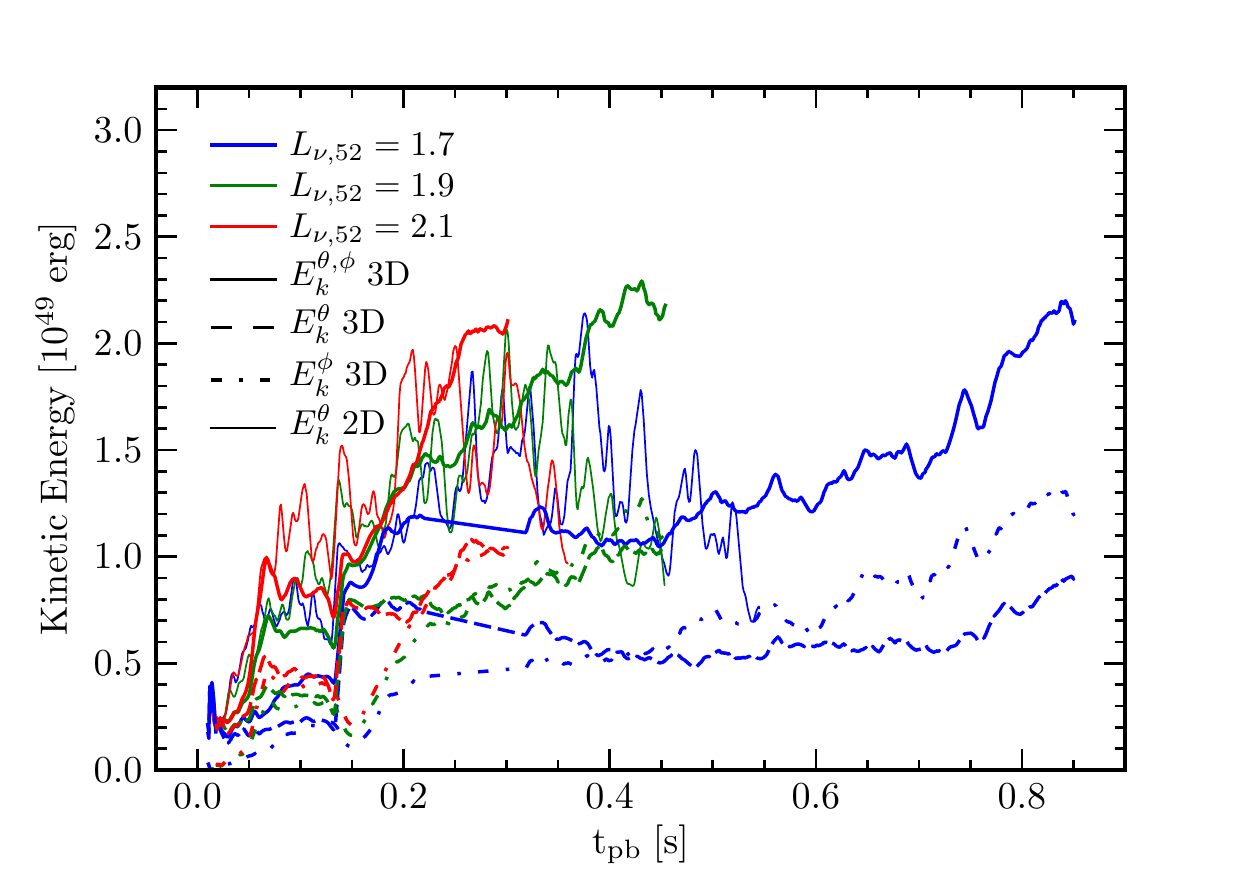}
\caption{Non-radial components of the kinetic energy in the gain
  region.  The transverse kinetic energy is higher in 2D than in 3D,
  but this is again a reflection of the greater mass in the gain
  region for 2D.}
\label{fig:ektht}
\end{figure}

Non-radial motion, particularly on large scales, has been suggested as
a primary factor resulting in easier explosions in multidimensional
simulations as compared with spherically-symmetric calculations
\citep{Hanke:2012dx}.  Such motion increases the matter dwell times in
the gain region and, thus, the net neutrino heating rate.  A measure
of non-radial motion is the gain region kinetic energy in transverse,
$\theta/\phi$-direction, motion, which we plot in Figure
\ref{fig:ektht}.  The kinetic energy of transverse motion in the gain
region is seen to corollate with increased neutrino luminosity.  The
transverse kinetic energy is also typically higher in 2D than in 3D.

Turbulence has been suggested to play an important role in the
supernova shock expansion in multidimensional simulations
\citep{Murphy:2011ci,Murphy:2013eg}.  The nature of turbulence between
2D and 3D, however, is fundamentally different.  The best-known
example of this fundamental difference is the so-called `inverse'
energy cascade in 2D: energy is transported to large scales in 2D
whereas in 3D energy is transported to smaller scales
\citep{Kraichnan:1967jk}.  The characteristic power-law slope of the
energy cascade is -5/3 in either the spherical harmonic mode, $\ell$,
or wavenumber, $k$.  Also in 2D, enstrophy, a quantity proportional to
the squared vorticity, is transported to smaller scales in a so-called
forward cascade with a characteristic power-law index of -3.

\begin{figure}
\centering
\begin{tabular}{cc}
\includegraphics[width=3.45in,trim= 0.05in 0in 0.4in 0.25in, clip]{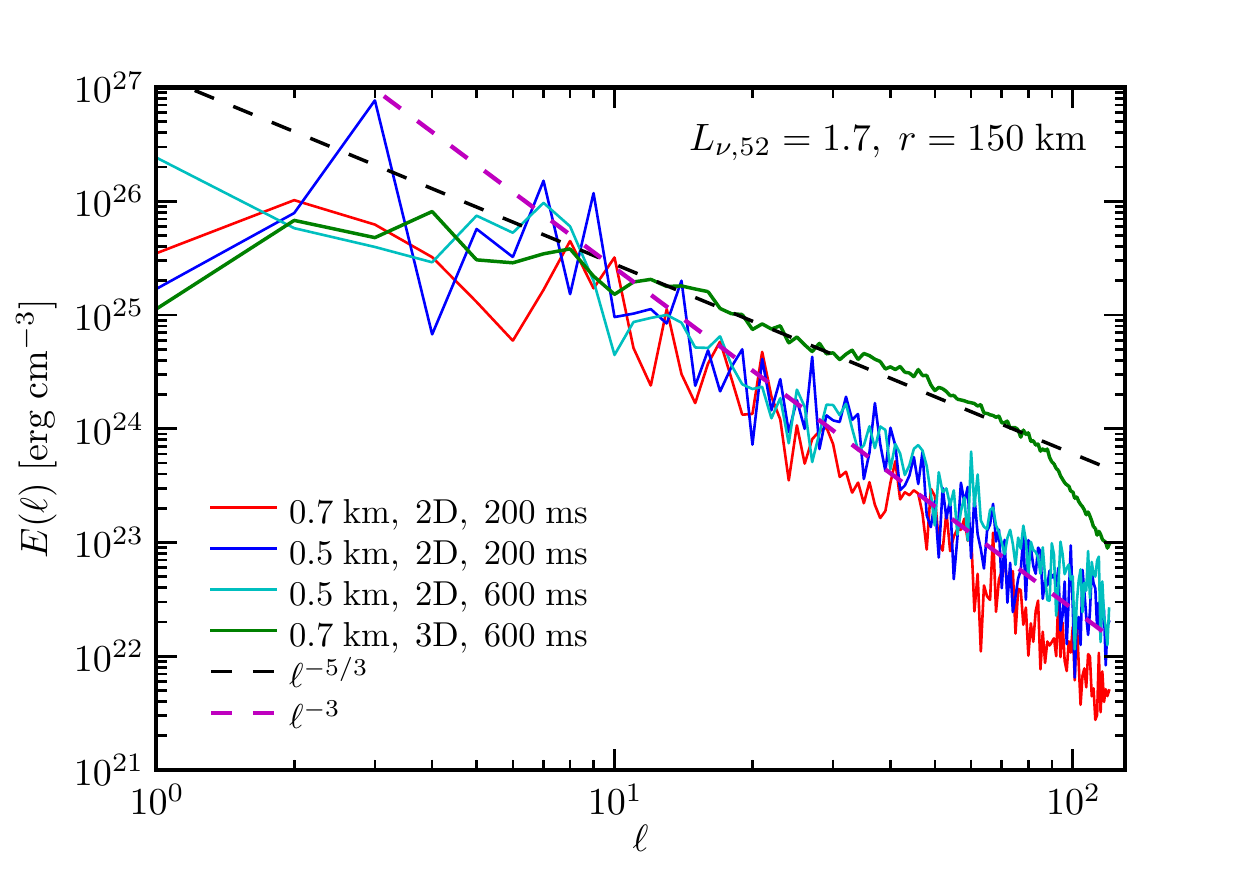}
\end{tabular}
\caption{Energy spectra of the $\theta$-direction kinetic energy, a
  proxy for the turbulent kinetic energy, in spherical harmonic basis
  calculated in a $\sim$10 km wide shell centered on 150 km.  We plot
  spectra for both 2D and 3D, and the spectra are averaged over 10 ms
  centered on the times indicated.  For 3D we find an inertial range
  from about $\ell=10-70$ over which the spectrum is roughly
  consistent with a $\ell^{-5/3}$ power-law.  Dissipation sets in
  around $\ell\sim70$ where the spectrum falls off more steeply than
  $\ell^{-5/3}$.  In all cases, the 2D spectra show more energy at
  $\ell=1$ than the 3D spectrum, reflective of the inverse energy
  cascade in 2D.  At around the driving scale of $\ell\sim10$, the 2D
  spectra follow a $\ell^{-3}$ power-law consistent with the forward
  enstrophy cascade.  The dissipation scale for the higher resolution
  2D simulations is at noticeably larger $\ell$.}
\label{fig:powSpec}
\end{figure}

\begin{figure*}
\centering
\begin{tabular}{cc}
\includegraphics[width=2.5in,trim= 2.25in 0in 1.25in 0in, clip]{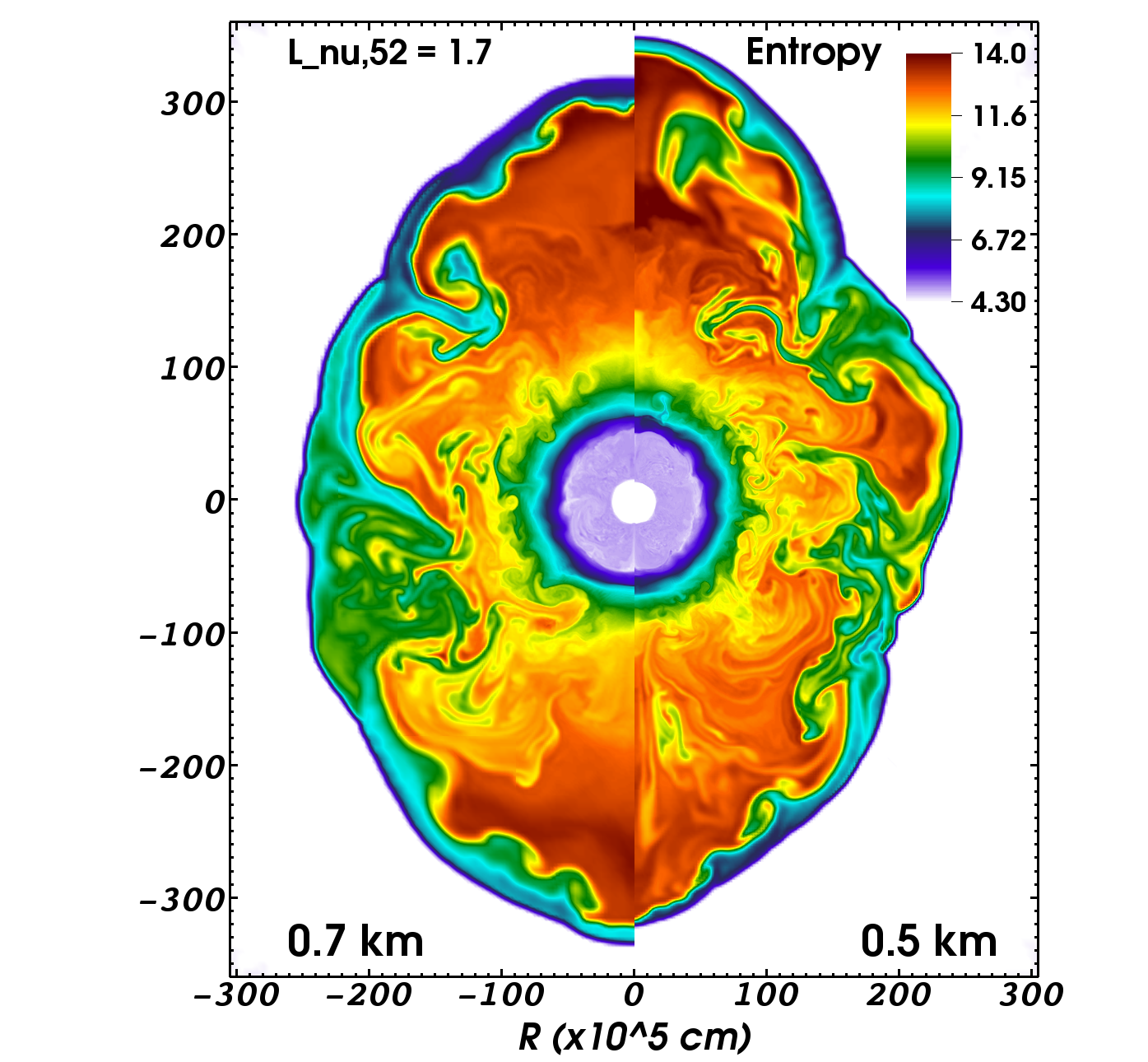}
\includegraphics[width=2.4in,trim= 0in 0in 4.25in 0in, clip]{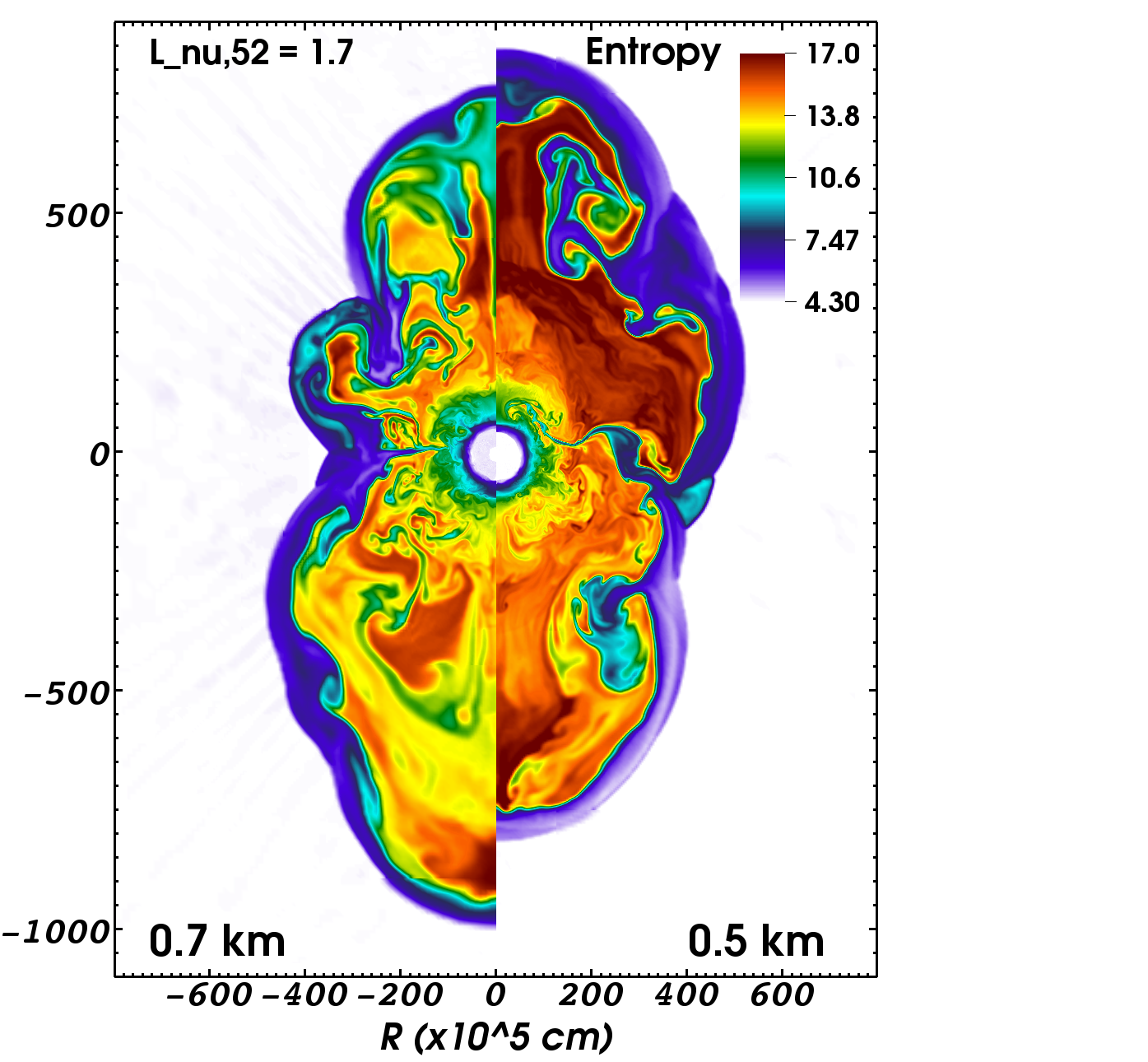}
\end{tabular}
\caption{Comparison between 2D simulations at different resolutions at
  200 ms (left) and time of explosion (right).  The higher resolution
  simulations are marked by deeper penetration of low-entropy down
  flows, increasing the surface are of buoyant plumes.}
\label{fig:comp2d}
\end{figure*}

In order to study the character of the turbulence in our simulations
we have computed the $\theta$-direction kinetic energy spectrum in
spherical harmonic basis where the energy in spherical harmonic mode
$\ell$ is 
\beq E(\ell) = \sum_{m=-\ell}^\ell a_{\ell m}^2, \eeq 
and the coefficients $a_{\ell m}$ are given by equation (\ref{eq:alm})
with $X(\theta,\phi) = \sqrt{\rho}(r,\theta,\phi)
v_{\theta}(r,\theta,\phi)$.  Transverse kinetic energy spectra for our
multidimensional simulations are shown in Figure \ref{fig:powSpec}
where we restrict the evaluation to a $\sim$10 km shell centered at
$r=150$ km.  We also average the spectra over 10 ms centered on the
respective labeled times.  The differing character of turbulence
between 2D and 3D is evident.  Beginning around $\ell=10$, which we
identify as the driving scale of the turbulence, the 3D spectrum is
consistent with a $\ell^{-5/3}$ power-law and the 2D spectra roughly
follow a $\ell^{-3}$ power-law, consistent with our expectations based
on turbulence theory \citep{Kolmogorov:1941vt,Kraichnan:1967jk}.  Also
evident is that there is much more energy at small scales (large
$\ell$'s) in 3D than in 2D, while 2D has more energy at large scales
($\ell =1-3$).  In a real sense, the different natures of these energy
spectra reflect the different characteristic buoyant plume sizes seen
in our simulations: plumes are smaller and more numerous in 3D as
compared with 2D.  This is then directly related to our argument that
the smaller plumes in 3D will experience a greater drag-to-buoyant
force, slowing their average ascent relative to the 2D case, resulting
in a less-rapid average shock expansion in 3D.

Figure \ref{fig:powSpec} also shows energy spectra for 2D at different
times and at different resolutions.  Note that for all the spectra
plotted the average shock radius is approximately the same at the time
shown (250-275 km, see Fig. \ref{fig:rshock}).  Considering the
evolution of the 2D energy spectra with time, the inverse energy
cascade is evident in the increase in energy by an order of magnitude
at $\ell=1$ while the spectrum in the inertial range, $\ell>10$,
remains relatively unchanged in the later time.  The lower-resolution
(0.7 km) 2D spectrum shows two important distinctions.  First,
dissipation sets in at smaller $\ell$, consistent with the assumption
that grid spacing determines the dissipation scale.  Second, at 200 ms
there is substantially more energy on large scales ($\ell=1,2$) than
for the 0.5 km resolution case.

It is of curious, perhaps surprising, significance that the energy
spectra seem to match the expected power-law slopes so well.  The
power-law scaling in the inertial ranges of -5/3 for 3D and -3 for 2D
are based on the assumption of fully-developed {\it isotropic}
turbulence \citep{Kolmogorov:1941vt,Kraichnan:1967jk}.  The turbulence
in our simulations is not isotropic but convective.  There is no model
of buoyant convective turbulence that would suggest any other
power-law scalings, but that they seem to be the same is an
interesting coincidence.

\section{Resolution Dependence}
\label{sec:resolution}

As discussed in the previous sections, we find that increasing the
resolution in 2D simulations results in later explosion times.  This
is an expected result of equation (\ref{eq:ratio}) that smaller
buoyant plumes rise more slowly than large buoyant plumes due to the
greater amount of drag force relative to buoyant force.  The different
convective plume size is evident by the differences in the energy
spectra in Figure \ref{fig:powSpec}, but also by visual inspection of
the 2D data.  In Figure \ref{fig:comp2d} we show a side-by-side
comparison of the 2D simulations with different resolution at two
different post-bounce times.  What is evident in this comparison is
that, even at 200 ms, the higher-resolution simulation shows a larger
number of low-entropy down flows and deep penetration of these down
flows.  The larger number of down flows breaks up the rising buoyant
plumes while their deeper penetration further increases the surface
are of the plumes and, therefore, the drag force felt by the plume.
This is especially evident at explosion time (right half of
Fig. \ref{fig:comp2d}).

This is simple to understand by considering the primary instabilities
that will contribute to break-up of the buoyant plumes.  The plumes
are inherently Rayleigh-Taylor unstable\footnote{In a sense, it is the
  Rayleigh-Taylor instability that drives the convection in the first
  place.} and will also develop parasitic Kelvin-Helmholtz
instabilities as they rise through the post-shock accretion flow.  In
the inviscid limit, the growth rate for both of these instabilities is
inversely proportional to the grid scale
\citep{Chandrasekhar:1961uk,Youngs:1984ds}, thus the higher-resolution
simulation experiences faster growth of these instabilities that
impede the rising plumes, slowing overall shock expansion.

\begin{figure}
\centering
\includegraphics[width=3.45in,trim= 0.05in 0in 0.4in 0.25in, clip]{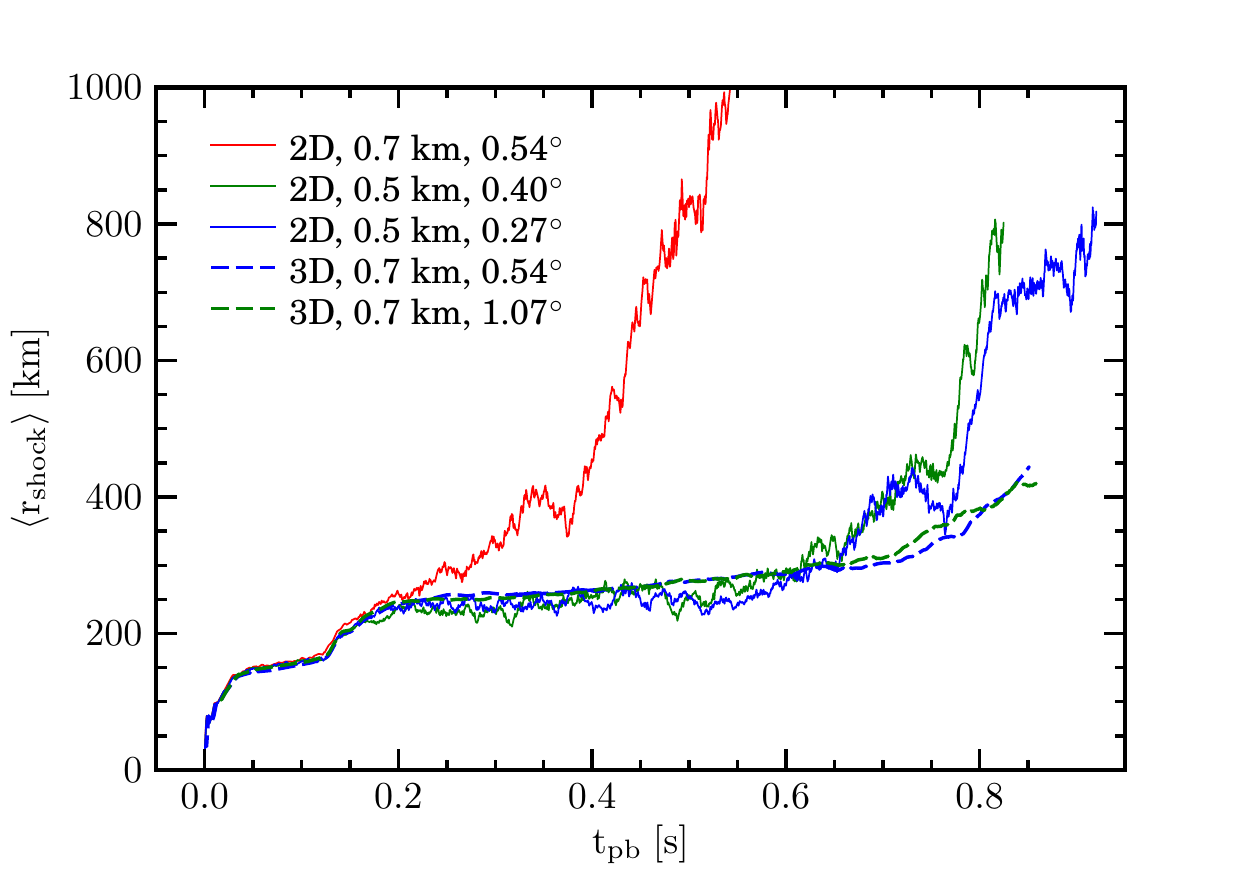}
\caption{Shock radius curves for 2D and 3D simulations with various
  resolutions.  The labels indicate maximum resolution as well as
  effective angular resolution for each simulation shown.  For 2D
  convergence with respect to shock radius history is obtained at a
  resolution of $0\fdg40$ while in 3D such convergence is already
  achieved at resolutions of $1\fdg07$. }
\label{fig:rshRes}
\end{figure}

\citet{Hanke:2012dx} report that increasing the resolution in 3D
results in delayed explosions while increasing resolution in 2D
results in earlier explosions.  Their 2D resolution study results are
contradictory to our simulation results.  Hanke et al.'s conclusion
that increasing resolution in 2D results in earlier explosions is
based on their simulations using only 400 radial zones.  When
considering their results with 600 or 800 radial zones, however, the
dependence on resolution is less clear; for some sets of 2D models
with more than 400 radial zones increasing the angular resolution
results in later explosions, as we have found.  Hanke et al. suggest
that this behavior with increased radial resolution is due to a
artificial density peak in the cooling region that grows with
increasing radial resolution, enhancing total cooling.  We do not see
such an artificial density peak in our simulations.  It is also worth
noting that our fiducial ``angular'' resolution of $0\fdg54$ is
substantially finer than the fiducial resolution considered by Hanke
et al. ($3\fdg$) and comparable to the finest resolution they use in
2D ($0\fdg5$).  \citet{Nordhaus:2010ct} and \citet{Dolence:2013iw} use
a fiducial resolution greater than ours (0.5 km) but do not conduct
resolution studies so it is unknown how their results would vary with
resolution.

In order to further investigate the resolution dependence of our
results, and whether they are ``converged'' with respect to grid
resolution, we have carried out additional simulations with varied
resolution.  In 2D we further increased the effective angular
resolution from $0\fdg4$ to $0\fdg27$ while keeping the maximum
resolution fixed at 0.5 km.  This moved the transition from $\Delta
x_i = 0.5$ km to $\Delta x_i = 1.0$ km from around a radius of 100 km
to 200 km.  In 3D, due to the great expense of higher resolution
simulations, we reduced the angular resolution from $0\fdg54$ to
$1\fdg07$ while keeping the maximum resolution fixed at 0.7 km.  This
moved the resolution transition from 0.5 km to 1.0 km from a radius of
100 km to around 50 km.  The shock radius curves for calculations
of various resolutions are shown in Figure \ref{fig:rshRes}.  All
these simulations use $L_{\nu_e,52} = 1.7$.  We see that in 2D
resolutions of $0\fdg4$ and $0\fdg27$ result in very similar shock
expansion histories and explosion times, indicating that the 2D results
are ``converged'' at an angular resolution of $0\fdg4$.  In 3D, the
$1\fdg07$ resolution simulation also results in a similar shock history
and explosion time as the $0\fdg54$ simulation.  This is encouraging
and interesting because it implies that the rate of convergence is
faster in 3D than in 2D.  Thus our fiducial 3D resolution is
sufficient to obtain convergence with respect to explosion time.

We note that convergence is a difficult notion in astrophysical
simulations that involve instabilities and turbulence such as ours.
Our simulations are not converged in the true sense that changing the
resolution would yield indistinguishable results.  For this to be the
case, we would need to reach numerical Reynolds numbers comparable to
the physical Reynolds numbers of the flow.  We are no where near
achieving this criterion in our calculations.  We have, however,
achieved convergence with resolution of the result we seek: the
explosion times.

\section{Discussion}
\label{sec:Discussion}

The results of previous studies similar to ours have been
contradictory.  The Princeton group has found that 3D simulations
explode sooner than 2D \citep{Nordhaus:2010ct,Dolence:2013iw} while
the Garching group finds little difference in the explosion times
between 3D and 2D \citep{Hanke:2012dx}.  The initial aspiration of our
study was to cast a tie-breaking vote in this discussion but instead
it has raised new questions by yielding a third result: our 3D
simulations explode {\it later} than our 2D simulations.  The source
of the disparity in the results from the different groups is still
unclear.  There are significant differences in numerical approach
between the Princton and Garching groups.  The Princeton group uses
the new code CASTRO \citep{Almgren:2010du} which implements a
directionally-unsplit piecewise parabolic method with the Riemann
solver of \citet{Bell:1989kq}.  CASTRO uses patch-based adaptive mesh
refinement \citep[c.f.,][]{Berger:1984js,Berger:1989gg} as provided by
the {\tt Boxlib} framework.  The Princeton simulations are run in 1D
spherical, 2D cylindrical, and 3D Cartesian coordinates.  The Garching
group uses the venerable hydrodynamics code PROMETHEUS
\citep{Fryxell:1991dh} which implements directionally-split PPM with a
`two shock' Riemann solver \citep{Colella:1985dc} in smooth flow and
the HLLE Riemann solver in shocks.  PROMETHEUS does not use AMR but
instead relies on 1D, 2D, and 3D spherical coordinates with
non-equidistant radial spacing.  Though perhaps the most important
differences between the approaches of the Princeton and Garching
groups are those dealing with approximations to the neutrino physics.
\citet{Nordhaus:2010ct} and \citet{Dolence:2013iw} follow closely the
approach of \citet{Murphy:2008ij}: deleptonization is approximated
using the density-dependent parameterization of
\citet{Liebendorfer:2005ft}, both pre- and post-bounce, and
post-bounce neutrino heating and cooling are considered locally based
on rates derived by \citet{Janka:2001fp}.  Heating and cooling are
shut off at high density by adding a $e^{-\tau_\nu}$ term to the
rates.  \citet{Hanke:2012dx} follow a very similar approach except for
the following.  They follow collapse and bounce to 15 ms post-bounce
with full neutrino transport in 1D and do not employ the
parameterization of \citet{Liebendorfer:2005ft} at all.  The
subsequent multidimensional evolution uses the same heating and
cooling rates as in \citet{Nordhaus:2010ct} but the neutrino optical
depth is computed by integrating the appropriate opacity in radius
whereas the Princeton group uses a density-dependent parameterization
for the optical depth \citep[J. Murphy, private communication, see
also][]{Couch:2013df}.  In their multidimensional post-bounce
calculations, the Garching group follows the evolution of the inner
core in 1D to avoid issues with grid convergence.

Our approach is very similar to those of Nordhaus et al. and Dolence
et al.: directionally-unsplit PPM in 1D spherical, 2D cylindrical, and
3D Cartesian with AMR.  We use the Liebendorfer parameterized
deleptonization pre- and post-bounce and approximate the neutrino
optical depth with a density-dependent piecewise fit, although we
scale optical depth differently than the Princeton group resulting in
a different normalization of the critical curves (see also the
discussion in Hanke et al.).  Notable differences in our scheme and
that of the Princeton group are we use an oct-tree, block-structured
AMR package that does not sub-cycle in time, i.e. all refinement
levels advance with the same time step size.  CASTRO uses adaptive
time refinement allowing coarser resolution levels to advance with
larger time steps.  And we use a ``hybrid'' Riemann solver that uses
the HLLC solver in smooth flows and the HLLE solver in shocks.
Another possibly significant difference amongst all three groups is in
the EOS implementation.  While each group is using the Shen et
al. high-density baryonic EOS for the referenced simulations, the
construction of the tables actually used is no doubt somewhat
different.  There could also be important differences in the details
of the implementations of the monopole gravity solver amongst the
different codes.  What is clearly mandated by the disparity in the
results is a rigorous code-to-code comparison.

Our results, particularly in 3D, support the conjecture that, for this
progenitor and treatment of neutrino physics, the SASI is subdominant
to neutrino-driven convection in advancing the average shock radius
\citep{Burrows:2012gc,Murphy:2013eg}. We are, however, very hesitant
to extrapolate this conclusion to all possible progenitors.  For
instance, \citet{Muller:2012kq} show clear evidence for a strong SASI
in the 2D explosion of a 27 $M_\sun$ progenitor using conformally-flat
general relativistic dynamics and full neutrino transport.  Although
recently \citet{Ott:2013gz} have simulated the same 27 $M_\sun$
progenitor with full 3D GR and a multispecies neutrino leakage scheme
and found that neutrino-driven convection becomes the dominant
instability in exploding models.

In the scenario that neutrino-driven convection dominates, the shock
expansion can be roughly described by buoyant plumes plumes rising
against the drag force exerted by the post-shock accretion flow
\citep{Dolence:2013iw}.  The buoyant force exerted by the neutrino
radiation on a plume will be proportional to the plumes subtended
solid angle times the optical depth of the plume multiplied by the
neutrino luminosity, i.e., the plume's volume.  The drag force exerted
by the accretion down flows will be proportional to the plume's
surface area.  A natural result of this model is that a single plume
with volume $V$ will rise more quickly than two plumes with volumes
$V/2$ due to the greater amount of total surface area, and hence drag
force, for the two plumes [see equation (\ref{eq:ratio})].  As
discussed at length in Section \ref{sec:interpret}, 3D simulations
develop smaller, more numerous buoyant plumes than 2D simulations.
Thus the balance between buoyancy and drag explains why the 3D models
explode later than the 2D models.  This picture also accounts for the
2D resolution dependence as higher-resolution will result in plumes
that have greater surface area-to-volume ratios.


%
%

\section{Conclusions}
\label{sec:Conclusions}

We have conducted a 1D, 2D, and 3D parameter study of neutrino-driven
core-collapse supernovae designed to explore the difference that 3D
makes on the explosion characteristics, particularly time of explosion
relative to 1D and 2D.  We find, as have a number of previous studies,
that the so-called `critical curve' in the neutrino luminosity-mass
accretion rate at explosion time plane is significantly lowered in
multiple dimensions relative to the spherical-symmetric case.  A novel
result of our study is we find that 3D explosions occur {\it later}
than 2D explosions at the same neutrino luminosity, i.e., the 3D
critical curve is {\it higher} than the 2D critical curve. We find
that the 2D results are resolution-dependent: increasing the
resolution in 2D delays explosion pushing the high-resolution 2D
critical curve very near to the 3D critical curve
(Fig. \ref{fig:mdot}).

We suggest that our results can be explained by the competition
between buoyancy and drag.  The shock expansion in our simulations is
dominated by the action of neutrino-driven buoyant convection
\citep[see also][]{Burrows:2012gc,Murphy:2013eg}.  In this case, the
shock expansion can be fit by the motion of a buoyant plume rising
through the shocked accretion flow \citep{Dolence:2013iw}.  The
buoyant force acting on the plume, provided by the absorbed neutrino
radiation, is proportional to a plume's volume whereas the drag force
resulting from the downward-flowing accretion is proportional to a
plume's area.  Thus, a plume's ascension velocity increases with
increasing volume-to-surface area ratio.  Rapid shock expansion,
therefore, is best abetted by plumes that subtend large solid angles.
We have shown that 3D simulations naturally result in many more,
smaller-solid angle plumes than comparable 2D simulations.  We posit
that this, then, is why our 3D simulations explode later than our 2D
simulations and why higher-resolution 2D simulations also result in
later explosions.

We examined several differences between our 2D and 3D simulations.  In
Section \ref{sec:shock} we explored the character of the shock motion
by calculating the first few spherical harmonic powers of the shock
deformation in 2D and 3D.  We find that the amplitudes of the
low-order, $\ell=1,2$, modes of the shock motion that are often
associated with the SASI are much reduced in 3D relative to 2D.  This
result is in qualitative agreement with \citet{Burrows:2012gc}.

When considering possible indicators of explosion, we find that the
``antesonic'' condition of \citet{Pejcha:2012cw} correlates well with
proximity to explosion.  We find that the mass-averaged entropy in the
gain region is typically higher in 3D than in 2D, but our results
indicate that is is not a good indicator of proximity to explosion, as
it was suggested to be by \citet{Nordhaus:2010ct}.  Instead this
higher average entropy, which is proportional to the time-integrated
neutrino heating per mass, only reflects that there is less mass in
the gain region in 3D as compared to 2D.  The greater gain region mass
in 2D also results in a greater net heating rate, which also aids
explosion relative to 3D.

The character of the non-radial motion and turbulence between our 2D
and 3D simulations is distinct.  In 3D, the forward energy cascade
transports energy to smaller and smaller scales until dissipation do
to finite grid size sets-in.  In 2D, the forward enstrophy cascade
results in a much less efficient transport of energy to small scales
while the inverse energy cascade actually transports energy from the
driving scales to larger scales.  This behavior is conducive to
explosion as it encourages both the growth of the low-order SASI and
the development of large buoyant plumes.

The cause of the differences in results among the groups attempting
similar multidimensional parameter studies \citep[e.g.,][ this
work]{Nordhaus:2010ct,Hanke:2012dx,Dolence:2013iw} remain to be
explained.  The simplified physics employed by such studies would make
a detailed code-to-code comparison effort far more straight-forward
than a comparison that included neutrino transport.  Such a
code-to-code comparison will hopefully be seen as a high priority and
be accomplished in the near future.

The approximations we employ are admittedly crude.  They have
tremendous merit, however, in that they facilitate 3D parameter
studies of CCSNe.  While certain conclusions about CCSNe based on
studies such as the present should be made with caution, the 3D
results appearing in the literature to-date make one undeniable point:
3D core-collapse supernovae are fundamentally and dramatically
different than 2D core-collapse supernovae.  The absence of
forced-symmetry in 3D makes an enormous impact on the character of the
shock motion and the development of neutrino-driven convection and
turbulence.  Our results indicate, however, that 3D alone may not be
the key to successful, robust explosions in more realistic
simulations.  Still, the enormous difference between 2D and 3D CCSN
simulations emphasizes and underlines the need for fully 3D
simulations by this community.

\acknowledgements 
The author thanks Evan O'Connor, Chris Daley, Carlo
Graziani, Cal Jordan, Jeremiah Murphy, and Christian Ott for helpful
and insightful conversations.  The author is especially grateful to
Todd Thompson for very valuable comments on the manuscript.  Support
for this work was provided by NASA through Hubble Fellowship grant
No. 51286.01 awarded by the Space Telescope Science Institute, which
is operated by the Association of Universities for Research in
Astronomy, Inc., for NASA, under contract NAS 5-26555.  The software
used in this work was in part developed by the DOE NNSA-ASC OASCR
Flash Center at the University of Chicago.  This research used
computational resources at ALCF at ANL, which is supported by the
Office of Science of the US Department of Energy under Contract
No. DE-AC02-06CH11357.  The author acknowledges the Texas Advanced
Computing Center (TACC) at The University of Texas at Austin for
providing high-performance computing, visualization, and data storage
resources that have contributed to the research results reported
within this paper

\bibliography{Bibliography}

\end{document}